\documentclass[11pt,english]{elsarticle}
\usepackage{mathptmx}
\usepackage[T1]{fontenc}
\usepackage[latin9]{inputenc}
\usepackage[letterpaper]{geometry}
\usepackage[active]{srcltx}
\usepackage{color}
\usepackage{bm}
\usepackage{amsmath}
\usepackage{amsthm}
\usepackage{graphicx}

\makeatletter
\numberwithin{equation}{section}
\numberwithin{figure}{section}

\DeclareMathAlphabet{\mathsfsl}{OT1}{cmss}{m}{sl}
\newcommand{\tensor}[1]{\stackrel{\leftrightarrow}{\mathsfsl{#1}}}
\renewcommand{\vec}[1]{\mathbf{#1}}

\newcommand{\vvectilde}{\widetilde{\vec{v}}}

\newcommand{\vparatilde}{{\widetilde v}_{||}}

\newcommand{\vperptilde}{{\widetilde v}_{\perp}}

\newcommand{\ftilde}{{\widetilde f}}

\makeatother

\usepackage{babel}
\begin{document}

\title{A conservative phase-space moving-grid strategy for a 1D-2V Vlasov-Fokker-Planck
Equation}

\author[lanl]{W. T. Taitano\corref{cor1}}

\ead{taitano@lanl.gov}

\author[lanl]{L. Chacón}

\author[lanl2]{A. N. Simakov}

\author[lanl]{S. E. Anderson}

\cortext[cor1]{Corresponding author}

\address[lanl]{Theoretical Division Los Alamos National Laboratory, Los Alamos,
NM 87545}

\address[lanl2]{Theoretical Design Division, Los Alamos National Laboratory, Los
Alamos, NM 87545}

\address{}
\begin{abstract}
We develop a conservative phase-space grid-adaptivity strategy for
the Vlasov-Fokker-Planck equation in a planar geometry. The velocity-space
grid is normalized to the thermal speed and shifted by the bulk-fluid
velocity. The configuration-space grid is moved according to a mesh-motion-partial-differential
equation (MMPDE), which equidistributes a monitor function that is
inversely proportional to the gradient-length scales of the macroscopic
plasma quantities. The grid adaptation ensures discrete conservation
of the collisional invariants (mass, momentum, and energy). The conservative
grid-adaptivity strategy provides an efficient scheme which resolves
important physical structures in the phase-space while controlling
the computational complexity at all times. We demonstrate the favorable
features of the proposed algorithm through a set of test cases of
increasing complexity.
\end{abstract}
\maketitle

\section{Introduction\label{sec:Introduction}}

The Vlasov-Fokker-Planck (VFP) collisional kinetic description, coupled
with Maxwell's equations, is regarded as a first-principles physical
model for describing weakly coupled plasmas in all collisionality
regimes, and accordingly, has a wide range of applications in laboratory
(e.g., magnetic and inertial thermonuclear fusion), space (e.g., Earth's
magnetosphere), and astrophysical (e.g., stellar mass ejections) plasmas.
In the VFP system, collisions are modeled by the Landau/Rosenbluth-Fokker-Planck
collision operator, which describes collisional relaxation of particle-velocity-distribution
functions in plasmas under the assumption of binary, grazing-angle
collisions \citep{rosenbluth,arsenev_1991conn_boltz_lfpe,desvillettes_1992_asymp_boltz_eqn_graze_col,degond_1992_fp_asymp_bolt_col_op_coul,goudon_1997_jsp,landau_1937_fokker_planck}. 

The system of VFP equations for various plasma species supports disparate
length and time scales. When coupled with strongly varying (in space
and time) plasma temperature, $T$, and bulk-flow velocity, $\vec{u}$,
this makes the system particularly challenging to solve with Eulerian
(grid-based) approaches. The challenges of temperature variation are
evident when one considers that the thermal speed, $v_{th}=\sqrt{2T/m}$,
provides a characteristic width of the distribution function and is
a function of the plasma temperature, $T$, and particle species mass,
$m$. In many practical applications of interest (e.g., inertial confinement
fusion {[}ICF{]}), $v_{th}$ variation for a given species can span
several orders of magnitude in both time and space. Also, mass disparities
result in strong $v_{th}$ separation for different species. Since
the velocity-space domain size is determined by the largest $v_{th}$,
and the velocity-space resolution by the smallest $v_{th}$, velocity-space
discretization with uniform Cartesian grids in such scenarios may
lead to impractical grid-size requirements. To complicate matters
further, a large velocity-space domain is required to accommodate
the bulk-flow separation in situations where cold, energetic beams
interact with a warm thermal background (i.e., $|\vec{u}|/v_{th}\gg1$),
further burdening the computational resource requirements. 

Several studies have recognized and tried to address these challenges
by normalizing the velocity coordinate to the local thermal speed
of the plasma and shifting by the bulk flow velocity \citep{larroche_2007_lsse_jcp,larroche_EPJ_2003_icf_fuel_ion_implosion_sim,jarema_CPC_2015_block_structured_grid_adaptive_vpace,peigney_JCP_2014_fp_kinetic_modeling_of_alpha}.
In this fashion, the grid expands/contracts as the plasma heats/cools,
and shifts as it accelerates/decelerates. Others have considered adaptivity
strategies based on either a combination of hierarchical grids \citep{Gutnic_CPC_2004_vlasov_phase_space_adaptivity,Gutnic_CPC_2006_vlasov_phase_space_adaptivity_wh_conservation}
by using multiresolution analysis techniques (very much like sparse-grid
techniques), or full adaptive-mesh-refinement (AMR) strategies in
phase-space \citep{kolobov_arxiv_2018_BFP_with_phase_space_amr},
providing promising paths for controlling the number of unknowns for
Eulerian and semi-Lagrangian schemes. Particularly relevant to this
study is the work in Ref. \citep{larroche_EPJ_2003_icf_fuel_ion_implosion_sim},
where the velocity-space domain was adapted for multiple ion species
based on a single local \emph{average} $v_{th}$ (over the ion species)
and hydrodynamic center-of-mass velocity of the plasma. This powerful
strategy enabled the first fully kinetic semi-Lagrangian implosion
simulations of inertial confinement fusion (ICF) capsules \citep{larroche_EPJ_2003_icf_fuel_ion_implosion_sim,larroche_pop_2012_Dhe_3_icf_sim,inglebert_epl_2014_species_separation_kinetic_effect_neutron_diagnostics},
but required \textcolor{black}{intermittent remapping in both the
configuration and velocity space}.  \textcolor{black}{Unfortunately,
all of the strategies outlined above do not conserve mass, momentum,
and energy (invariants of the VFP equation) and some also break the
structured nature of the grid.}

Recently, a novel strategy was proposed in Ref. \citep{Taitano_2018_vrfp_1d2v_implicit}
for a system of 1D-2V Vlasov-Fokker-Planck equations  that deals with
strong temperature disparity (in both space and time), avoids remapping,
and works on structured meshes. The strategy employs a multiple-grid
approach by normalizing each species' velocity to its local and instantaneous
thermal speed. The Vlasov-Fokker-Planck equations were transformed
analytically and then discretized on a mesh. The transformed equations
retained the continuum conservation symmetries, which were then enforced
in the discrete via additional nonlinear constraints. This strategy
ensures that the species' distribution functions are always well resolved
regardless of temperature or mass disparity. However, this strategy
does not allow for species with strongly disparate bulk-flow velocities,
ultimately requiring a large number of mesh points in scenarios when
$|\vec{u}|/v_{th}\gg1$. Additionally, the evolution of structures
in the configuration space was not taken into account, requiring a
large number of unknowns to track and resolve sharp features, such
as shocks.

In this study, we extend the conservative, multiple-dynamic velocity-space
adaptivity strategy in a 1D-2V Cartesian system developed in Ref.
\citep{Taitano_2018_vrfp_1d2v_implicit} to incorporate the further
velocity and configuration space (i.e., phase-space) adaptivity discussed
above. We consider a quasi-neutral plasma with multiple kinetic ion
species and fluid electrons. As before, ionic species are evolved
on a velocity-space grid normalized to a temporally and spatially
varying characteristic speed, $v^{*}$ (a function of their $v_{th}$),
but also \emph{shifted }by their characteristic velocity, $\vec{u}^{*}$
(a function of their $\vec{u}$). Also, we evolve the configuration-space
coordinate on a logically Cartesian grid, and the Jacobian of the
transformation is evolved using a mesh motion partial differential
equation (MMPDE) \citep{huang_1994_jcp_mmpde_ep,huang_1994_siam_J_numer_anal_mmpdes_ep,budd_huang_russell_2009}.
The approach relies on the equidistribution of monitor functions,
which in turn are defined as functions of local gradient-length scales
of the plasma. This analytical transformation of the VFP equation
introduces additional inertial terms, which are carefully discretized
to ensure simultaneous conservation of mass, momentum, and energy. 

The rest of the paper is organized as follows. Section \ref{sec:VRFP+electrons}
introduces the ion-VFP and fluid-electron equations and discusses
their conservation properties. In Sec. \ref{sec:coordinate_transformation},
we introduce the normalized Vlasov-Fokker-Planck equation and detail
the coordinate transformation. In Sec. \ref{sec:MMPDE}, we briefly
describe our MMPDE approach, the choice of monitor function, and an
algorithm to evolve the configuration-space grid. In Sec. \ref{sec:Numerical-implementation},
we provide a detailed discussion on the implementation of the proposed
scheme in the following order: 1) a discretization of the Vlasov-Fokker-Planck
equation with the additional inertial terms, 2) a discretization of
the fluid electron temperature equation,  3) a discrete-conservation
strategy for the Vlasov component with the added inertial terms, 4)
the discretization of our MMPDE equation and the configuration-space
grid velocity, and 5) spatial-temporal evolution strategy for $v^{*}$
and $\vec{u}^{*}$. The numerical performance of the scheme is demonstrated
with various multi-species tests of varying degrees of complexity
in Sec. \ref{sec:numerical-results}. Finally, we conclude in Sec.
\ref{sec:conclusions}.

\section{The System of Vlasov Fokker-Planck ion and fluid electrons equations
\label{sec:VRFP+electrons}}

A dynamic evolution of weakly-coupled collisional plasmas can be described
by a system of Vlasov-Fokker-Planck equation for species $\alpha$
distribution functions (PDF), $f_{\alpha}\left(\vec{\vec{x},\vec{v},t}\right)$,
in configuration space, $\vec{x}$, velocity space, $\vec{v}$, and
time, $t$: 
\begin{equation}
\partial_{t}f_{\alpha}+\nabla_{x}\cdot\left(\vec{v}f_{\alpha}\right)+\frac{q_{\alpha}}{m_{\alpha}}\frac{\partial}{\partial\vec{v}}\cdot\left[\left(\vec{E}+\vec{v}\times\vec{B}\right)f_{\alpha}\right]=\sum_{\beta=1}^{N_{s}}C_{\alpha\beta},
\end{equation}
where $\vec{E}$ is the electric field, $\vec{B}$ is the magnetic
field, $N_{s}$ is the total number of plasma species, and $C_{\alpha\beta}$
is the Fokker-Planck collision operator for species $\alpha$ colliding
with species $\beta$: 

\begin{equation}
C_{\alpha\beta}=\Gamma_{\alpha\beta}\frac{\partial}{\partial\vec{v}}\cdot\left[\tensor{D}_{\beta}\cdot\nabla_{v}f_{\alpha}-\frac{m_{\alpha}}{m_{\beta}}\vec{{A}}_{\beta}f_{\alpha}\right].\label{eq:fokker_planck_operator}
\end{equation}
Here, $\Gamma_{\alpha\beta}=\frac{2\pi Z_{\alpha}^{2}Z_{\beta}^{2}e^{4}\Lambda_{\alpha\beta}}{m_{\alpha}^{2}},$
$\tensor D{}_{\beta}$ and $\vec{{A}}_{\beta}$ are the tensor-diffusion
and friction coefficients for species $\beta$, $m_{\alpha}$ and
$m_{\beta}$ are the masses of species $\alpha$ and $\beta$, respectively,
$Z_{\alpha}=q_{\alpha}/e$ is the ionization state of species $\alpha$,
$e$ is the proton charge, and $\Lambda_{\alpha\beta}$ is the Coulomb
logarithm (unless otherwise specified, $\Lambda_{\alpha\beta}=10$
is assumed for simplicity in this study for all species). In this
work, we adopt the Rosenbluth potential formulation \citep{rosenbluth}
to compute $\tensor D_{\beta}$ and $\vec{A}_{\beta}$ and refer the
readers to Refs. \citep{Taitano_2018_vrfp_1d2v_implicit,Taitano_2015_rfp_0d2v_implicit,Taitano_2016_rfp_0d2v_implicit}
for the detailed numerical treatment of the collision operator.

The collision operator, Eq. (\ref{eq:fokker_planck_operator}), preserves
the positivity of $f_{\alpha}$, and conserves mass, momentum, and
energy. The conservation properties stem from the following symmetries
\citep{braginskii}:
\begin{eqnarray}
\left<1,C_{\alpha\beta}\right>_{\vec{v}} & = & 0,\label{eq:charge-cons}\\
m_{\alpha}\left<\vec{v},C_{\alpha\beta}\right>_{\vec{v}} & = & -m_{\beta}\left<\vec{v},C_{\beta\alpha}\right>_{\vec{v}},\label{eq:momentum-cons}\\
m_{\alpha}\left<\frac{v^{2}}{2},C_{\alpha\beta}\right>_{\vec{v}} & = & -m_{\beta}\left<\frac{v^{2}}{2},C_{\beta\alpha}\right>_{\vec{v}},\label{eq:energy-cons}
\end{eqnarray}
where the inner product is defined as $\left<A,B\right>_{\vec{v}}=2\pi\int_{-\infty}^{\infty}dv_{||}\int_{0}^{\infty}dv_{\perp}v_{\perp}A(\vec{v})\,B(\vec{v})$
(for the cylindrically symmetric coordinate system in the velocity
space employed herein). These conservation symmetries can be enforced
in the discrete following the general procedures discussed in Refs.
\citep{Taitano_2015_rfp_0d2v_implicit,Taitano_2016_rfp_0d2v_implicit}.

In this study, we consider a 1D planar geometry in the configuration
space without a magnetic field. Without loss of generality and similarly
to the previous study \citep{Taitano_2018_vrfp_1d2v_implicit}, we
consider a 2V cylindrically symmetric coordinate system in the velocity
space. We adopt a fluid-electron model and employ the quasi-neutrality
and ambipolarity approximations. We obtain the following simplified
system of equations comprised of the ion Vlasov-Fokker-Planck equation
(per species $\alpha$),

\begin{equation}
\partial_{t}f_{\alpha}+\partial_{x}\left(v_{||}f_{\alpha}\right)+\frac{q_{\alpha}}{m_{\alpha}}E_{||}\partial_{v_{||}}f_{\alpha}=\sum_{\beta}^{N_{s}}C_{\alpha\beta}+C_{\alpha e}\label{eq:1d_vfp_eqn}
\end{equation}
and the electron-temperature equation,

\begin{equation}
\frac{3}{2}\frac{\partial}{\partial t}\left[n_{e}T_{e}\right]+\frac{5}{2}\partial_{x}\left[u_{||,e}n_{e}T_{e}\right]+\partial_{x}Q_{||,e}-q_{e}n_{e}u_{||,e}E_{||}=\sum_{\alpha}^{N_{s}}W_{e\alpha}.\label{eq:1d_electron_equation}
\end{equation}
Here, $n_{e}=\sum_{\alpha}^{Ns}q_{\alpha}n_{\alpha}$ is the electron
number density, $u_{||,e}=\sum_{\alpha}^{N_{s}}q_{\alpha}n_{\alpha}u_{||,\alpha}/n_{e}$
is the parallel electron fluid velocity, and $E_{||}$ is the parallel
electric field, evaluated using the Ohm's law. Also, $Q_{||,e}$ is
the parallel electron-heat flux, $T_{e}$ is the electron temperature,
$W_{e\alpha}$ describes the electron-ion energy exchange, and the
detailed theoretical and numerical treatment of these quantities can
be found in \citep{simakov_PoP_2014_e_transp_wh_multi_ion} and \citep{Taitano_2018_vrfp_1d2v_implicit},
respectively. The electron-ion collision operator $C_{\alpha e}$
in Eq. (\ref{eq:1d_vfp_eqn}) is given by:

\begin{equation}
C_{\alpha e}=\Gamma_{\alpha e}\nabla_{v}\cdot\left[\tensor D_{\alpha e}\cdot\nabla_{v}f_{\alpha}-\frac{m_{\alpha}}{m_{e}}\vec{A}_{\alpha e}f_{\alpha}\right],\label{eq:ion_electron_collision_operator}
\end{equation}
where we adopt the reduced ion-electron potentials \citep{hazeltine_1991_plas_confin}
(for the full details on the numerics of ensuring discrete conservation,
refer to Ref. \citep{Taitano_2018_vrfp_1d2v_implicit}).

\section{Coordinate transformations of the ion Vlasov-Fokker-Planck and fluid-electron
equations\label{sec:coordinate_transformation}}

\begin{figure}[h]
\begin{centering}
\includegraphics[height=5cm]{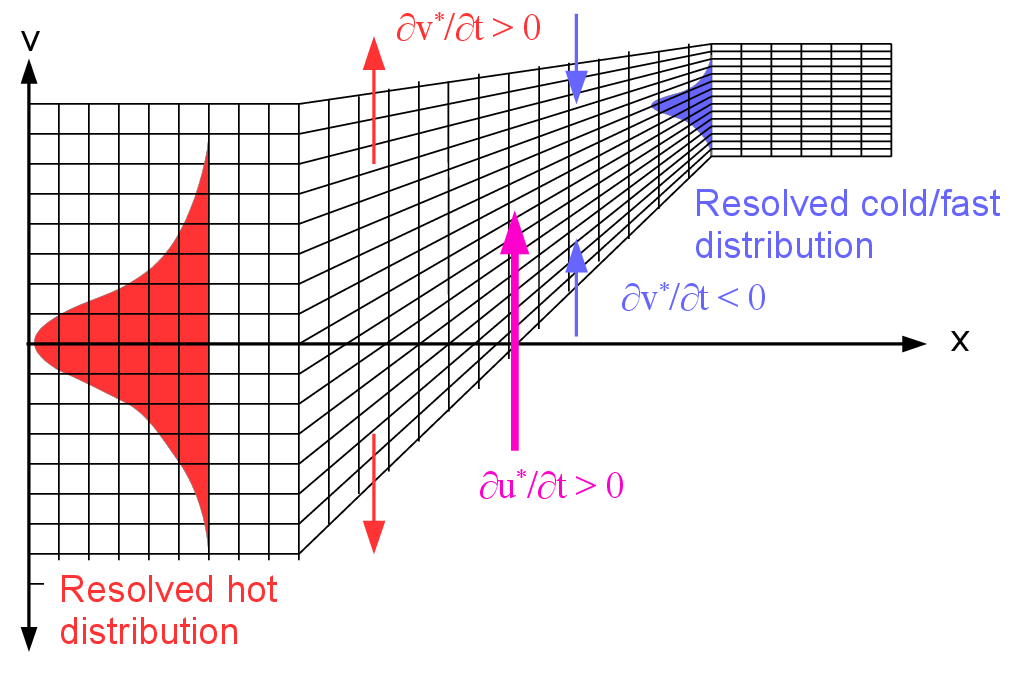}
\par\end{centering}
\caption{Illustration of the velocity space adaptivity.\label{fig:vth-disparity-in-space-wh-adaptivity-1}}
\end{figure}

We transform the velocity space for a species $\alpha$ by normalizing
to a speed, $v_{\alpha}^{*}\left(x,t\right)$, and shifting by a velocity,
$u_{||,\alpha}^{*}\left(x,t\right)$, (related to their thermal speed
and bulk flow velocity, respectively) as follows:

\[
\vec{\widetilde{v}}_{\alpha}=\frac{\vec{v}}{v_{\alpha}^{*}}-\widehat{u}_{||,\alpha}^{*}\vec{e}_{||},\;\frac{\partial}{\partial\widetilde{\vec{v}}_{\alpha}}=v_{\alpha}^{*}\frac{\partial}{\partial\vec{v}_{\alpha}},\;\widetilde{f}_{\alpha}=\left(v_{\alpha}^{*}\right)^{3}f_{\alpha}.
\]
Here, the hat denotes quantities normalized to $v_{\alpha}^{*}$,
the tilde denotes quantities additionally shifted by $\widehat{u}_{||,\alpha}^{*}$,
and $\vec{e}_{||}=\left[1,0,0\right]^{T}$ is the unit vector parallel
to the flow direction. To simplify notation, from here on we will
denote $\vec{v}_{\alpha}$ with $\vec{v}$, i.e., we will omit the
species subscript. As an example, the density, drift, and temperature
moments are defined as:

\[
n_{\alpha}=\left\langle 1,\widetilde{f}_{\alpha}\right\rangle _{\widetilde{\vec{v}}},\;u_{||,\alpha}=v_{\alpha}^{*}\widehat{u}_{||,\alpha}=v_{\alpha}^{*}\left\langle \widetilde{v}_{||}+\widehat{u}_{||,\alpha}^{*},\widetilde{f}_{\alpha}\right\rangle _{\widetilde{\vec{v}}}/n_{\alpha},\;T_{\alpha}=\left(v_{\alpha}^{*}\right)^{2}\widetilde{T}_{\alpha}=\frac{m_{\alpha}\left(v_{\alpha}^{*}\right)^{2}\left\langle \left(\widetilde{\vec{v}}+\widehat{u}_{||,\alpha}^{*}-\widehat{u}_{||,\alpha}\right)^{2},\ftilde\right\rangle _{\vvectilde}}{3\left\langle 1,\ftilde\right\rangle _{\vvectilde}}
\]
where $\left\langle \left(\cdot\right),\ftilde_{\alpha}\right\rangle _{\vvectilde}=2\pi\int_{-\infty}^{\infty}d\vparatilde\int_{0}^{\infty}\left(\cdot\right)\ftilde_{\alpha}\vperptilde d\vperptilde$.
The normalization of other relevant quantities and the collision operator
with respect to $v_{\alpha}^{*}$ are discussed in Ref. \citep{Taitano_2016_rfp_0d2v_implicit}.
We note that, as $v_{\alpha}^{*}$ is a function of local $v_{th}$,
and $\widehat{u}_{||}^{*}$ a function of the local $u_{||}$ for
a given plasma species (elaborated in Sec. \ref{subsec:Evolution-equation-of-vstar}),
the grid will expand/contract as the plasma heats/cools, and translates
as the plasma accelerates/decelerates. For an illustration of this
process, refer to Fig. \ref{fig:vth-disparity-in-space-wh-adaptivity-1}.

In the configuration space, we perform a coordinate transformation,
$x=x\left(\xi,t\right)$, to evolve the dependent quantities in the
logically Cartesian computational space, $\xi$. The Vlasov-Fokker-Planck
equation is transformed into the new coordinate system, $\left(\xi,\widetilde{v}_{||},\widetilde{v}_{\perp}\right)$,
as follows (details of a derivation are provided in \ref{app:coordinate_transformation_details}):
\begin{eqnarray}
\left.\frac{\partial\left(J_{\xi}\widetilde{f}_{\alpha}\right)}{\partial t}\right|_{\xi,\widetilde{\vec{v}}}+\frac{\partial}{\partial\xi}\left[\left(v_{\alpha}^{*}\left[\widetilde{v}_{||}+\widehat{u}_{||,\alpha}^{*}\right]-\dot{x}\right)\widetilde{f}_{\alpha}\right]_{\widetilde{\vec{v}},t}-\frac{J_{\xi}}{v_{\alpha}^{*}}\frac{\partial}{\partial\widetilde{\vec{v}}}\cdot\left[\frac{\partial}{\partial t}\left(v_{\alpha}^{*}\left[\vec{\widetilde{v}}+\widehat{u}_{||,\alpha}^{*}\vec{e}_{||}\right]\right)\widetilde{f}_{\alpha}\right]_{\xi,t}\nonumber \\
-\frac{1}{v_{\alpha}^{*}}\frac{\partial}{\partial\widetilde{\vec{v}}}\cdot\left[\frac{\partial}{\partial\xi}\left(v_{\alpha}^{*}\left[\vec{\widetilde{v}}+\widehat{u}_{||,\alpha}^{*}\vec{e}_{||}\right]\right)\left(v_{\alpha}^{*}\left[\widetilde{v}_{||}+\widehat{u}_{||,\alpha}^{*}\right]-\dot{x}\right)\widetilde{f}_{\alpha}\right]_{\xi,t}+J_{\xi}\frac{q_{\alpha}}{m_{\alpha}}\frac{E_{||}}{v_{\alpha}^{*}}\left.\frac{\partial\widetilde{f}_{\alpha}}{\partial\widetilde{v}_{||}}\right|_{\xi,t}=J_{\xi}\left[\sum_{\beta}\widetilde{C}_{\alpha\beta}+\widetilde{C}_{\alpha e}\right].\label{eq:vfp_eqn_transformed}
\end{eqnarray}
Here, $J_{\xi}=\partial_{\xi}x$ is the Jacobian of transformation
in the configuration space and $\dot{x}=\partial_{t}x$ is the grid
speed. Similarly, the fluid electron temperature equation is given
in the transformed coordinate system by:

\begin{equation}
\frac{3}{2}\frac{\partial}{\partial t}\left(J_{\xi}n_{e}T_{e}\right)_{\xi}+\frac{5}{2}\frac{\partial}{\partial\xi}\left(u_{||,e}n_{e}T_{e}\right)_{t}-\frac{3}{2}\frac{\partial}{\partial\xi}\left(\dot{x}n_{e}T_{e}\right)_{t}+\left.\frac{\partial Q_{||,e}}{\partial\xi}\right|_{t}-q_{e}n_{e}u_{||,e}E_{||}J_{\xi}=J_{\xi}\sum_{\alpha}^{N_{sp}}W_{e\alpha}.\label{eq:1d_fluid_electron_transformed}
\end{equation}

\section{Moving Mesh Partial Differential Equation (MMPDE)\label{sec:MMPDE}}

To allow the desired grid adaptivity in the configuration space, we
employ a moving-mesh-partial-differential equation (MMPDE) strategy
\citep{huang_1994_jcp_mmpde_ep,huang_1994_siam_J_numer_anal_mmpdes_ep,li_1998_siam_jci_comp_stability_of_mmpde,budd_huang_russell_2009}.
In particular, we choose the MMPDE5 scheme of Ref. \citep{huang_1994_jcp_mmpde_ep},
\begin{equation}
\partial_{t}x=\tau_{x}^{-1}\partial_{\xi}\left[\omega_{x}\partial_{\xi}x\right],\label{eq:mmpde_equation}
\end{equation}
with boundary conditions, $\left.x\right|_{\xi=0}=x_{min}$ and $\left.x\right|_{\xi=1}=x_{max}$.
Here $\tau_{x}$ is the equilibration time-scale in which the grid
reaches an optimal distribution for a given monitor function, $\omega_{x}$
(i.e., for $\tau_{x}\to0$, we recover $\partial_{\xi}\left[\omega_{x}J_{\xi}\right]=0$),
and acts as a temporal smoothing mechanism for the grid evolution.
The quality of the grid is governed by the choice of $\omega_{x}$.
We choose it to depend on inverse gradient-length scales of $n$,
$v_{th}$, and $u_{||}$, and $T_{e}$, and define as:

\begin{equation}
\omega_{x}=\sqrt{\frac{1}{2}\sum_{\alpha}^{N_{sp}}\left[\left(L_{n,\alpha}^{-1}\right)^{2}+\left(L_{v_{th},\alpha}^{-1}\right)^{2}+\left(L_{u_{||},\alpha}^{-1}\right)^{2}\right]+\frac{1}{2}\left(L_{T_{e}}^{-1}\right)^{2}},\label{eq:inverse_grad_scale_length_monitor_function}
\end{equation}
where, in the discrete, the gradient-length scales are computed as

\begin{eqnarray}
L_{n,\alpha}^{-1}=\left|\frac{\delta_{\xi}n_{\alpha}}{n_{\alpha}}\right|+\delta_{min,n},\;\;\;\;L_{v_{th},\alpha}^{-1}=\left|\frac{\delta_{\xi}v_{th,\alpha}}{v_{th,\alpha}}\right|+\delta_{min,v_{th}},\label{eq:inverse_gradient_length_scale}\\
L_{u_{||},\alpha}^{-1}=\left|\frac{\delta_{\xi}u_{||,\alpha}}{v_{th,\alpha}}\right|+\delta_{min,u_{||}},\;\;\;\;L_{T_{e}}^{-1}=\left|\frac{\delta_{\xi}T_{e}}{T_{e}}\right|+\delta_{min,T_{e}}.\nonumber 
\end{eqnarray}
Here, $\delta_{\xi}{\cal M}=0.5\left[{\cal M}\left(\xi+\epsilon_{\xi},t\right)-{\cal M}\left(\xi-\epsilon_{\xi},t\right)\right]$
denotes a finite-differencing operation in the $\xi$ coordinate,
where $\epsilon_{\xi}$ is the finite difference factor about point
$\xi$, and $\delta_{min,{\cal M}}$ is the floor for the relative
variation of the quantity ${\cal M}$. In practical calculations,
$\delta_{min,{\cal M}}$ may be individually specified in order to
weight the variations of moments separately (e.g., $v_{th}$ and $u_{||}$
variations can be weighted more than variation of $n$ to ensure a
smooth grid in velocity space {[}to be discussed shortly{]}). In this
study, we choose $\delta_{min,{\cal M}}=0.025$ for all ${\cal M}$.
Note that, since $u_{||,\alpha}$ is a vector quantity (and therefore
can vanish), we normalize it with respect to $v_{th}$. In addition,
to avoid an arbitrarily fine mesh, we limit the minimum-to-maximum
ratio of the monitor function to a cutoff value, $\eta$, by modifying
the monitor function by a constant, $a$,

\begin{equation}
\omega_{x}\leftarrow\omega_{x}+a\;\;\;\textnormal{if}\;\;\;\frac{\omega_{x,min}}{\omega_{x,max}}\le\eta,\label{eq:flooring_of_monitor_function}
\end{equation}
where $a=\left(\eta\omega_{x,max}-\omega_{x,min}\right)/\left(1-\eta\right)$
and otherwise stated, $\eta=10^{-2}$ is used in this study. Further,
for the stability of the mesh-motion scheme, one must specify a value
for $\tau_{x}$ that captures the macroscopic flow evolution well
enough while ensuring reasonable grid velocities. Unless otherwise
stated, we choose $\tau_{x}=0.5$ in this study, but it can be problem
dependent.

\section{Numerical Implementation\label{sec:Numerical-implementation}}

\subsection{Discretization of the VFP equation with inertial terms\label{subsec:discretization_of_the_vfp_equation_wh_adaptivity}}

Following Refs. \citep{Taitano_2018_vrfp_1d2v_implicit}, we discretize
VFP equation, Eq. (\ref{eq:vfp_eqn_transformed}), using finite volumes
in a 1D planar logical space, $\xi$, and 2V transformed cylindrical-velocity
space with azimuthal symmetry, ($\widetilde{v}_{||},\widetilde{v}_{\perp}$).
We compute the discrete volume for cell \emph{i,j,k} (corresponding
to logical, parallel, and perpendicular velocity space coordinates,
respectively) as:
\[
\Delta V_{i,j,k}=J_{\xi,i}\Delta\xi\Delta V_{j,k},
\]
with
\[
\Delta V_{j,k}=2\pi\widetilde{v}_{\perp,k}\Delta\widetilde{v}_{||}\Delta\widetilde{v}_{\perp},
\]
where $J_{\xi,i}=\frac{x_{i+1/2}-x_{i-1/2}}{\Delta\xi}$ is the discrete
Jacobian in the configuration space, and $\Delta\xi$, $\Delta\vparatilde$,
and $\Delta\vperptilde$ are the mesh spacings in the logical and
the transformed parallel- and perpendicular-velocity spaces, respectively.
For uniform logical and normalized velocity-space meshes, we have:

\[
\Delta\xi=\frac{1}{N_{\xi}},\,\,\Delta\vparatilde=\frac{\widetilde{L}_{||}}{N_{||}},\,\,\Delta\vperptilde=\frac{\widetilde{L}_{\perp}}{N_{\perp}},
\]
where, $\widetilde{L}_{||}$ and $\widetilde{L}_{\perp}$ are the
normalized parallel and perpendicular velocity-space domain sizes,
respectively; and $N_{\xi}$, $N_{||}$, and $N_{\perp}$ are the
corresponding numbers of cells. The mesh is arranged such that cell
faces map to the domain boundary (and therefore outermost cell centers
are half a mesh-spacing away from the boundary). We define the distribution
function $f$ at cell centers. 

Velocity-space inner products are approximated via a mid-point quadrature
rule as
\begin{equation}
\left<A,B\right>_{\widetilde{\vec{v}}}\approx\left<A,B\right>_{\delta\widetilde{\vec{v}}}=\sum_{j=1}^{N_{||}}\sum_{k=1}^{N_{\perp}}A_{j,k}B{}_{j,k}\Delta V_{j,k}\label{eq:fv_quadrature}
\end{equation}
for scalars, and
\begin{eqnarray}
\left<\vec{A},\vec{B}\right>_{\widetilde{\vec{v}}}\approx\left<\vec{A},\vec{B}\right>_{\delta\widetilde{\vec{v}}}=\nonumber \\
2\pi\left[\sum_{j=0}^{N_{||}}\sum_{k=1}^{N_{\perp}}\widetilde{v}_{\perp,k}\Delta\widetilde{v}_{||}\Delta\widetilde{v}_{\perp}A_{||,j+1/2,k}B{}_{||,j+1/2,k}+\sum_{j=1}^{N_{||}}\sum_{k=0}^{N_{\perp}}\widetilde{v}_{\perp,k+1/2}\Delta\widetilde{v}_{||}\Delta\widetilde{v}_{\perp}A_{\perp,j,k+1/2}B{}_{\perp,j,k+1/2}\right]
\end{eqnarray}
for vectors (with their components at cell faces denoted by the half-integer
indices $j+1/2$, $k+1/2$).

We discretize Eq. (\ref{eq:vfp_eqn_transformed}) in a conservative
form as

\begin{eqnarray}
\frac{c^{(p+1)}J_{\xi,i}^{(p+1)}\widetilde{f}_{\alpha,i,j,k}^{(p+1)}+c^{(p)}J_{\xi,i}^{(p)}\widetilde{f}_{\alpha,i,j,k}^{(p)}+c^{(p-1)}J_{\xi,i}^{(p-1)}\widetilde{f}_{\alpha,i,j,k}^{(p-1)}}{\Delta t^{(p)}}\nonumber \\
+\underbrace{\frac{F_{x,\alpha,i+1/2,j,k}^{(p+1)}-F_{x,\alpha,i-1/2,j,k}^{(p+1)}}{\Delta\xi}}_{\textcircled a}+\underbrace{\frac{F_{\dot{x},\alpha,i+1/2,j,k}^{(p+1)}-F_{\dot{x},\alpha,i-1/2,j,k}^{(p+1)}}{\Delta\xi}}_{\textcircled b}+\underbrace{\frac{J_{\alpha,acc,i,j+1/2,k}^{(p+1)}-J_{\alpha,acc,i,j-1/2,k}^{(p+1)}}{\Delta\widetilde{v}_{||}}}_{\textcircled c}\\
+\underbrace{\left[\frac{J_{||,t,\alpha,i,j+1/2,k}^{(p+1)}-J_{||,t,\alpha,i,j-1/2,k}^{(p+1)}}{\Delta\widetilde{v}_{||}}+\frac{\widetilde{v}_{\perp,k+1/2}J_{\perp,t,\alpha,i,j,k+1/2}^{(p+1)}-\widetilde{v}_{\perp,k-1/2}J_{\perp,t,\alpha,i,j,k-1/2}^{(p+1)}}{\widetilde{v}_{\perp,k}\Delta\widetilde{v}_{\perp}}\right]}_{\textcircled d}\nonumber \\
+\underbrace{\left[\frac{J_{||,x,\alpha,i,j+1/2,k}^{(p+1)}-J_{||,x,\alpha,i,j-1/2,k}^{(p+1)}}{\Delta\widetilde{v}_{||}}+\frac{\widetilde{v}_{\perp,k+1/2}J_{\perp,x,\alpha,i,j,k+1/2}^{(p+1)}-\widetilde{v}_{\perp,k-1/2}J_{\perp,x,\alpha,i,j,k-1/2}^{(p+1)}}{\widetilde{v}_{\perp,k}\Delta\widetilde{v}_{\perp}}\right]}_{\textcircled e}\nonumber \\
=J_{\xi,i}^{(p+1)}\left[\sum_{\beta}^{N_{s}}\left.\widetilde{C}_{\alpha\beta}^{(p+1)}\right|_{i,j,k}+\left.\widetilde{C}_{\alpha e}^{(p+1)}\right|_{i,j,k}\right].\label{eq:vfp_discrete_eqn}
\end{eqnarray}
Here, $c^{(p+1)}$, $c^{(p)}$, and $c^{(p-1)}$ are the coefficients
for the second-order backwards difference formula (BDF2) \citep{byrne-acmtms-75-bdf}
and superscripts $(p)$ is the discrete time index. 

The term $\textcircled a$ corresponds to the discretization of the
spatial streaming term, with

\begin{equation}
F_{x,\alpha,i+1/2,j,k}^{(p+1)}=v_{\alpha,i+1/2}^{*,(p)}\left(\widetilde{v}_{||,j}+\widehat{u}_{||,\alpha,i+1/2}^{*,(p)}\right)\textnormal{Interp}\left(\widetilde{v}_{||,j}+\widehat{u}_{||,\alpha,i+1/2}^{*,(p)},\widetilde{f}_{\alpha}^{(p+1)}\right)_{i+1/2,j,k},\label{eq:discrete_streaming_flux}
\end{equation}

\[
v_{\alpha,i+1/2}^{*,(p)}=\frac{v_{\alpha,i+1}^{*,(p)}+v_{\alpha,i}^{*,(p)}}{2},\;\textnormal{and}\;\widehat{u}_{||,\alpha,i+1/2}^{*,(p)}=\frac{\widehat{u}_{||,\alpha,i}^{*,(p)}+\widehat{u}_{||,\alpha,i+1}^{*,(p)}}{2},
\]
where $\textnormal{Interp}\left(a,\phi\right)_{face}$ is an advection
interpolation operator of a scalar $\phi$ at a cell face with a given
velocity $a$, which can be written in general as
\begin{equation}
\textnormal{Interp}\left(a,\phi\right)_{face}=\sum_{i'=1}^{N}\omega_{face,i'}\left(a,\phi\right)\phi_{i'}.\label{eq:face_interpolation_rule}
\end{equation}
The coefficients $\omega_{face,i'}$ are the interpolation weights
for the spatial cells $i'$ surrounding the cell face of interest.
In this study, they are determined by the SMART discretization \citep{smart}\textcolor{black}{,
which ensures the positivity of the solution and is well-posed for
nonlinear iterative methods. }

The term $\textcircled b$ corresponds to the inertial term in the
configuration space, arising from the moving grid, with

\begin{equation}
F_{\dot{x},\alpha,i+1/2,j,k}^{(p+1)}=-\dot{x}_{i+1/2}^{(p+1)}\textnormal{Interp}\left(-\dot{x}_{i+1/2}^{(p+1)},\widetilde{f}_{\alpha}^{(p+1)}\right)_{i+1/2,j,k},\label{eq:discrete_grid_motion_flux}
\end{equation}
where the definition of $\dot{x}_{i+1/2}$ is given in Sec. \ref{subsec:grid_velocity_discretization}. 

The term $\textcircled c$ corresponds to the electrostatic-acceleration
term with

\begin{equation}
J_{\alpha,acc,i,j+1/2,k}^{(p+1)}=J_{\xi}\frac{q_{\alpha}}{m_{\alpha}}\frac{E_{||,i}^{(p+1)}}{v_{\alpha,i}^{*,p}}\textnormal{Interp}\left(q_{\alpha}E_{||,i}^{(p+1)},\widetilde{f}_{\alpha}^{(p+1)}\right)_{i,j+1/2,k}.\label{eq:discrete_acceleration_flux}
\end{equation}

The term $\textcircled d$ corresponds to the inertial terms due to
temporal variation of the velocity space metrics (i.e., $v_{\alpha}^{*}$
and $\widehat{u}_{||,\alpha}^{*}$) with

\begin{equation}
J_{||,t,\alpha,i,j+1/2,k}^{(p+1)}=\gamma_{t,\alpha,i,j+1/2,k}^{(p+1)}{\cal I}_{||,t,\alpha,i,j+1/2,k}^{(p)}\textnormal{Interp}\left({\cal I}_{||,t,\alpha,i+1/2,j+1/2,k}^{(p+1)},\widetilde{f}_{\alpha}^{(p+1)}\right)_{i,j+1/2,k}\label{eq:discrete_temporal_flux_parallel}
\end{equation}
and

\begin{equation}
J_{\perp,t,\alpha,i,j,k+1/2}^{(p+1)}=\gamma_{t,\alpha,i,j,k+1/2}^{(p+1)}{\cal I}_{\perp,t,\alpha,i,j,k+1/2}^{(p+1)}\textnormal{Interp}\left({\cal I}_{\perp,t,\alpha,i,j,k+1/2}^{(p+1)},\widetilde{f}_{\alpha}^{(p+1)}\right)_{i,j,k+1/2},\label{eq:discrete_temporal_flux_perpendicular}
\end{equation}
where $\gamma_{t,\alpha}=\gamma_{t,\alpha}\left(\xi,\widetilde{\vec{v}}\right)$
is the discrete nonlinear constraint function that enforces the simultaneous
momentum- and energy-conservation symmetries arising from the temporal
variations of the metrics (to be discussed shortly),

\begin{eqnarray}
{\cal I}_{||,t,\alpha,i,j+1/2,k}^{(p+1)}=-\left[\frac{J_{\xi,i}^{(p+1)}}{v_{\alpha,i}^{*,(p)}}\frac{\partial}{\partial t}\left(v_{\alpha}^{*}\left[\widetilde{v}_{||,j+1/2}+\widehat{u}_{||,\alpha}^{*}\right]\right)_{i}^{(p+1)}\right]\approx\nonumber \\
-\frac{J_{\xi,i}^{(p+1)}}{v_{\alpha,i}^{*,(p)}\Delta t^{(p)}}\left[c^{(p+1)}v_{\alpha,i}^{*,(p)}\left(\widetilde{v}_{||,j+1/2}+\widehat{u}_{||,\alpha,i}^{*,(p)}\right)+c^{(p)}v_{\alpha,i}^{*,(p-1)}\left(\widetilde{v}_{||,j+1/2}+\widehat{u}_{||,\alpha,i}^{*,(p-1)}\right)\right.\label{eq:temporal_inertial_coeff_parallel}\\
\left.+c^{(p-1)}v_{\alpha,i}^{*,(p-2)}\left(\widetilde{v}_{||,j+1/2}+\widehat{u}_{||,\alpha,i}^{*,(p-2)}\right)\right]
\end{eqnarray}
and
\begin{equation}
{\cal I}_{\perp,t,\alpha,i,j,k+1/2}^{(p+1)}=-\frac{J_{\xi,i}^{(p+1)}}{v_{\alpha,i}^{*,(p)}}\left[\frac{\partial}{\partial t}\left(v_{\alpha}^{*}\widetilde{v}_{\perp}\right)_{i,k+1/2}^{(p+1)}\right]\approx-\frac{J_{\xi,i}^{(p+1)}}{v_{\alpha,i}^{*,(p)}}\left[\frac{c^{(p+1)}v_{\alpha,i}^{*,(p)}+c^{(p)}v_{\alpha,i}^{*,(p-1)}+c^{(p-1)}v_{\alpha,i}^{*,(p-2)}}{\Delta t^{(p)}}\widetilde{v}_{\perp,k+1/2}\right].\label{eq:temporal_inertial_coeff_perpendicular}
\end{equation}
As described in Ref. \citep{Taitano_2016_rfp_0d2v_implicit}, we lag
the time level between the BDF2 coefficients and the normalization
speed (and similarly, the shift velocity) to avoid over-constraining
the nonlinear residual (see the reference for further detail). 

The term $\textcircled e$ corresponds to the inertial terms due to
the spatial variation of the metrics with

\begin{equation}
J_{||,x,\alpha,i,j+1/2,k}^{(p+1)}=0.5\left[J_{||,x,\alpha,i+1/2,j+1/2,k}^{(p+1),-}+J_{||,x,\alpha,i-1/2,j+1/2,k}^{(p+1),+}\right]\label{eq:discrete_spatial_inertial_flux_parallel}
\end{equation}

\begin{equation}
J_{\perp,x,\alpha,i,j,k+1/2}^{(p+1)}=0.5\left[J_{\perp,x,\alpha,i+1/2,j,k+1/2}^{(p+1),-}+J_{\perp,x,\alpha,i-1/2,j,k+1/2}^{(p+1),+}\right],\label{eq:discrete_spatial_inertial_flux_perpendicular}
\end{equation}
where

\begin{equation}
J_{||,x,\alpha,i\pm1/2,j+1/2,k}^{(p+1),\mp}=\gamma_{x,i\pm1/2,j+1/2,k}^{(p+1)}{\cal I}_{||,x,\alpha,i\pm1/2,j+1/2,k}^{(p+1),\mp}\textnormal{Interp}\left({\cal I}_{||,x,\alpha,i\pm1/2,j+1/2,k}^{(p+1),\mp},\widetilde{f}_{\alpha}^{(p+1)}\right)_{i,j+1/2,k}\label{eq:discrete_spatial_inertial_flux_paralle_ipm}
\end{equation}
and
\begin{equation}
J_{\perp,x,\alpha,i\pm1/2,j,k+1/2}^{(p+1),\mp}=\gamma_{x,i\pm1/2,j,k+1/2}^{(p+1)}{\cal I}_{\perp,x,\alpha,i\pm1/2,j,k+1/2}^{(p+1),\mp}\textnormal{Interp}\left({\cal I}_{\perp,x,\alpha,i\pm1/2,j,k+1/2}^{(p+1),\mp},\widetilde{f}_{\alpha}^{(p+1)}\right)_{i,j,k+1/2}.\label{eq:discrete_spatial_inertial_flux_perpendicular_ipm}
\end{equation}
Here, similarly to $\gamma_{t}$, $\gamma_{x}=\gamma_{x}\left(\xi,\widetilde{\vec{v}}\right)$
is the discrete nonlinear constraint function which enforces the simultaneous
mass, momentum, and energy conservation symmetries arising from the
coordinate transformation,

\begin{eqnarray}
{\cal I}_{||,x,\alpha,i+1/2,j+1/2,k}^{(p+1),\mp}=-\frac{1}{v_{\alpha,i+1/2}^{*,\mp}}\left\{ \left[v_{\alpha}^{*}\left(\widetilde{v}_{||}+\widehat{u}_{||,\alpha}^{*}\right)-\dot{x}\right]\frac{\partial}{\partial\xi}\left[v_{\alpha}^{*}\left(\widetilde{v}_{||}+\widehat{u}_{||,\alpha}^{*}\right)\right]\right\} _{i+1/2,j+1/2,k}^{(p+1)}\nonumber \\
\approx-\frac{v_{\alpha,i+1/2}^{*,(p)}\left(\widetilde{v}_{||,j+1/2}+\widehat{u}_{||,\alpha,i+1/2}^{*,(p)}\right)-\dot{x}_{i+1/2}^{(p+1)}}{v_{\alpha,i+1/2\mp1/2}^{*,(p)}}\frac{v_{\alpha,i+1}^{*,(p)}\left(\widetilde{v}_{||,j+1/2}+\widehat{u}_{||,\alpha,i+1}^{*,(p)}\right)-v_{\alpha,i}^{*,(p)}\left(\widetilde{v}_{||,j+1/2}+\widehat{u}_{||,\alpha,i}^{*,(p)}\right)}{\Delta\xi}\label{eq:discrete_spatial_inertial_flux_parallel_coefficient}
\end{eqnarray}
and

\begin{eqnarray}
{\cal I}_{\perp,x,\alpha,i+1/2,j,k+1/2}^{(p),\mp}=-\frac{1}{v_{\alpha,i+1/2\mp1/2}^{*}}\left\{ \left[v_{\alpha}^{*}\left(\widetilde{v}_{||}+\widehat{u}_{||,\alpha}^{*}\right)-\dot{x}\right]\frac{\partial}{\partial\xi}\left(v_{\alpha}^{*}\widetilde{v}_{\perp}\right)\right\} _{i+1/2,j,k+1/2}^{(p+1)}\nonumber \\
\approx-\frac{v_{\alpha,i+1/2}^{*,(p)}\left(\widetilde{v}_{||,j}+\widehat{u}_{||,\alpha,i+1/2}^{*,(p)}\right)-\dot{x}_{i+1/2}^{(p+1)}}{v_{\alpha,i+1/2\mp1/2}^{*,(p)}}\frac{v_{\alpha,i+1}^{*,(p)}\widetilde{v}_{\perp,k+1/2}-v_{\alpha,i}^{*,(p)}\widetilde{v}_{\perp,k+1/2}}{\Delta\xi}.\label{eq:discrete_spatial_inertial_flux_perpendicular_coefficient}
\end{eqnarray}

Finally, the right-hand-side of Eq. (\ref{eq:vfp_discrete_eqn}) corresponds
to the Fokker-Planck collision operator and its treatment is discussed\textcolor{black}{{}
in detail in Refs. \citep{Taitano_2015_rfp_0d2v_implicit,Taitano_2016_rfp_0d2v_implicit}.
In this study, we use the Du Toit-O'Brien-Vann (DOV) differencing
scheme \citep{DOV_CPC_effective_advection_tensor_diffusion} for the
tensor diffusion operator in the collision }term\textcolor{black}{{}
to cast the off-diagonal component of this operator as a nonlinear
advection term. This strategy allows the use of many high-order, maximum-principle-preserving
discretization schemes (e.g., SMART) and simplifies the preconditioning
strategy (i.e., upwind differencing can be used for this term).}

\subsection{Discretization of the Fluid Electron Equation\label{subsec:discretization_of_fluid_electron_equation}}

The electron temperature equation in the transformed coordinate system,
Eq. (\ref{eq:1d_fluid_electron_transformed}), is discretized using
a finite-volume scheme in space and BDF2 in time:
\begin{eqnarray}
\frac{3}{2}\frac{c^{(p+1)}J_{\xi,i}^{(p+1)}n_{e,i}^{(p+1)}T_{e,i}^{(p+1)}+c^{(p)}J_{\xi,i}^{(p)}n_{e,i}^{(p)}T_{e,i}^{(p)}+c^{(p-1)}J_{\xi,i}^{(p-1)}n_{e,i}^{(p-1)}T_{e,i}^{(p-1)}}{\Delta t^{(p)}}+\nonumber \\
\frac{5}{2}\frac{u_{||,e,i+1/2}^{(p+1)}\left(\widetilde{n_{e}T_{e}}\right)_{i+1/2}^{(p+1)}-u_{||,e,i-1/2}^{(p+1)}\left(\widetilde{n_{e}T_{e}}\right)_{i-1/2}^{(p+1)}}{\Delta\xi}-\frac{3}{2}\frac{\dot{x}_{i+1/2}^{(p+1)}\left(\widetilde{n_{e}T_{e}}\right)_{i+1/2}^{(p+1)}-\dot{x}_{i-1/2}^{(p+1)}\left(\widetilde{n_{e}T_{e}}\right)_{i-1/2}^{(p+1)}}{\Delta\xi}+\\
\frac{Q_{||,e,i+1/2}^{(p+1)}-Q_{||,e,i-1/2}^{(p+1)}}{\Delta\xi}-q_{e}\left[n_{e}u_{||,e}E_{||}\right]_{i}^{(p+1)}J_{\xi,i}^{(p+1)}=\label{eq:discrete_electron_eqn}\\
\left[3\nu_{e\alpha,i}^{(p+1)}\frac{m_{e}}{m_{\alpha}}n_{e,i}^{(p+1)}\left(T_{\alpha,i}^{(p+1)}-T_{e,i}^{(p+1)}\right)-F_{||,\alpha e,i}^{(p+1)}u_{||,\alpha,i}^{(p+1)}\right]J_{\xi,i}^{(p+1)}.
\end{eqnarray}
Here, the tilde denotes a cell-face discretization for the advection
quantities (SMART in this study). The Joule-heating quantity (last
term on the left-hand-side) is defined so as to enforce conservation
properties in the discrete with the detailed expression as well as
treatment of all other terms provided in Ref. \citep{Taitano_2018_vrfp_1d2v_implicit}. 

\subsection{Discretization of the Vlasov component: exact conservation properties}

This section describes the procedure to ensure the set of exact conservation
symmetries of the Vlasov piece in the ion kinetic equation in the
presence of phase-space grid adaptivity. In this, we follow a procedure
similar to that discussed in Refs. \citep{Taitano_2016_rfp_0d2v_implicit,Taitano_2018_vrfp_1d2v_implicit},
but with significant simplification in the derivation of the symmetries.
We note that the conservation properties associated with the temporal
and spatial variation of metrics can be shown independently. We begin
by developing discretizations for fully mass, momentum, and energy
conservation in spatially homogeneous system (0D), and then a spatially
inhomegenous case (1D) in a periodic spatial domain \textit{without}
any background field (conservation with electric field can be shown
separately, as shown in Ref. \citep{Taitano_2018_vrfp_1d2v_implicit}). 

\subsubsection{Temporal variation of $v^{*}$ and $\widehat{u}_{||}^{*}$}

If we consider only the terms associated with the temporally varying
metrics in Eq. (\ref{eq:vfp_eqn_transformed}) (i.e., a homogeneous
plasma), we obtain the following simplified form of the Vlasov equation:

\begin{equation}
\partial_{t}\left(J_{\xi}\widetilde{f}\right)_{\xi,\widetilde{\vec{v}}}-\frac{J_{\xi}}{v^{*}}\frac{\partial}{\partial\widetilde{\vec{v}}}\cdot\left(\left.\partial_{t}\vec{v}\right|_{\xi,\widetilde{\vec{v}}}\widetilde{f}\right)_{\xi,t}=0,\label{eq:simplified_temporal_only_vlasov}
\end{equation}
where $\partial_{t}\left.\vec{v}\right|_{\xi,\widetilde{\vec{v}}}=\partial_{t}\left[v^{*}\left(\vec{\widetilde{v}}+\widehat{u}_{||}^{*}\vec{e}_{||}\right)\right]_{\xi,\widetilde{\vec{v}}}$.
In the continuum, mass conservation is defined as:
\begin{equation}
\int_{0}^{1}d\xi\left\langle m,\partial_{t}\left(J_{\xi}\widetilde{f}\right)\right\rangle _{\widetilde{\vec{v}}}=0.
\end{equation}
This can be shown trivially due to the divergence form of the inertial
terms.

Momentum conservation is defined as
\begin{equation}
\int_{0}^{1}d\xi\left\langle m,\partial_{t}\left(v_{||}J_{\xi}\widetilde{f}\right)\right\rangle _{\widetilde{\vec{v}}}=0.
\end{equation}
This can be shown (for brevity, dropping the explicit integration
in $\xi$) by multiplying Eq. (\ref{eq:simplified_temporal_only_vlasov})
by $mv_{||}$, and using the chain rule to obtain:
\[
m\left\{ \partial_{t}\left(v_{||}J_{\xi}\ftilde\right)_{\xi,\widetilde{\vec{v}}}-\left[\underbrace{J_{\xi}\ftilde\partial_{t}\left.v_{||}\right|_{\xi,\widetilde{v}_{||}}}_{\textcircled a}+\underbrace{\frac{J_{\xi}v_{||}}{v^{*}}\frac{\partial}{\partial\widetilde{\vec{v}}}\cdot\left(\left.\partial_{t}\vec{v}\right|_{\xi,\widetilde{\vec{v}}}\widetilde{f}\right)_{\xi,t}}_{\textcircled b}\right]\right\} =0.
\]
Taking $\left\langle 1,\left(\cdot\right)\right\rangle _{\widetilde{\vec{v}}}$
of above, the terms $\textcircled a$ and $\textcircled b$ yields:
\[
\left\langle 1,\textcircled a\right\rangle _{\widetilde{\vec{v}}}=J_{\xi}\left\langle \left.\partial_{t}v_{||}\right|_{\xi,\widetilde{v}_{||}},\widetilde{f}\right\rangle _{\widetilde{\vec{v}}},
\]
\[
\left\langle 1,\textcircled b\right\rangle _{\widetilde{\vec{v}}}=J_{\xi}\left\langle \frac{v_{||}}{v^{*}},\frac{\partial}{\partial\vec{\widetilde{v}}}\cdot\left(\left.\partial_{t}\vec{v}\right|_{\xi,\widetilde{\vec{v}}}\widetilde{f}\right)_{\xi,t}\right\rangle _{\widetilde{\vec{v}}}=-J_{\xi}\left\langle \left.\partial_{t}v_{||}\right|_{\xi,\widetilde{v}_{||}},\widetilde{f}\right\rangle _{\widetilde{\vec{v}}},
\]
canceling each other and leaving $\left\langle m,\partial_{t}\left(v_{||}J_{\xi}\widetilde{f}\right)_{\xi,\widetilde{\vec{v}}}\right\rangle _{\widetilde{\vec{v}}}=0$.
Thus, the key is to ensure that in the discrete we satisfy the following
relation:

\begin{equation}
\left\langle v_{||},\partial_{t}\left(J_{\xi}\widetilde{f}\right)_{\xi,\widetilde{\vec{v}}}-\frac{J_{\xi}}{v^{*}}\frac{\partial}{\partial\widetilde{\vec{v}}}\cdot\left(\left.\partial_{t}\vec{v}\right|_{\xi,\widetilde{\vec{v}}}\widetilde{f}\right)_{\xi,t}\right\rangle _{\vec{\widetilde{v}}}=\left\langle 1,\partial_{t}\left(v_{||}J_{\xi}\widetilde{f}\right)_{\xi,\widetilde{\vec{v}}}\right\rangle _{\vec{\widetilde{v}}}.\label{eq:temporal_continuum_momentum_constraint}
\end{equation}

Similarly, energy conservation is defined as

\begin{equation}
\int_{0}^{1}d\xi\left\langle m,\partial_{t}\left(\frac{v^{2}}{2}J_{\xi}\widetilde{f}\right)\right\rangle _{\widetilde{\vec{v}}}=0.
\end{equation}
This can be shown by multiplying Eq. (\ref{eq:simplified_temporal_only_vlasov})
by $m\frac{v^{2}}{2}$, and using the chain rule to obtain:
\[
m\left\{ \partial_{t}\left(\frac{v^{2}}{2}J_{\xi}\ftilde\right)_{\xi,\widetilde{\vec{v}}}-\left[\underbrace{J_{\xi}\ftilde\partial_{t}\left.\frac{v^{2}}{2}\right|_{\xi,\widetilde{\vec{v}}}}_{\textcircled a}+\underbrace{\frac{J_{\xi}v^{2}}{2v^{*}}\frac{\partial}{\partial\widetilde{\vec{v}}}\cdot\left(\left.\partial_{t}\vec{v}\right|_{\xi,\widetilde{\vec{v}}}\widetilde{f}\right)_{\xi,t}}_{\textcircled b}\right]\right\} =0.
\]
Taking $\left\langle 1,\left(\cdot\right)\right\rangle _{\widetilde{\vec{v}}}$
of above, the terms $\textcircled a$ and $\textcircled b$ yields:
\[
\left\langle 1,\textcircled a\right\rangle _{\widetilde{\vec{v}}}=J_{\xi}\left\langle \left.\partial_{t}\frac{v^{2}}{2}\right|_{\xi,\widetilde{\vec{v}}},\widetilde{f}\right\rangle _{\widetilde{\vec{v}}},
\]
\[
\left\langle 1,\textcircled b\right\rangle _{\widetilde{\vec{v}}}=J_{\xi}\left\langle \frac{v^{2}}{2v^{*}},\frac{\partial}{\partial\vec{\widetilde{v}}}\cdot\left(\left.\partial_{t}\vec{v}\right|_{\xi,\widetilde{\vec{v}}}\widetilde{f}\right)_{\xi,t}\right\rangle _{\widetilde{\vec{v}}}=-J_{\xi}\left\langle \vec{v},\left.\partial_{t}\vec{v}\right|_{\xi,\widetilde{\vec{v}}}\widetilde{f}\right\rangle _{\widetilde{\vec{v}}}=-J_{\xi}\left\langle \left.\partial_{t}\frac{v^{2}}{2}\right|_{\xi,\widetilde{\vec{v}}},\widetilde{f}\right\rangle _{\widetilde{\vec{v}}},
\]
canceling each other and leaving $\left\langle m,\partial_{t}\left(\frac{v^{2}}{2}J_{\xi}\widetilde{f}\right)_{\xi,\widetilde{\vec{v}}}\right\rangle _{\widetilde{\vec{v}}}=0$.
Thus, the key is to ensure that in the discrete, we satisfy the following
relation:

\begin{equation}
\left\langle \frac{v^{2}}{2},\partial_{t}\left(J_{\xi}\widetilde{f}\right)_{\xi,\widetilde{\vec{v}}}-\frac{J_{\xi}}{v^{*}}\frac{\partial}{\partial\widetilde{\vec{v}}}\cdot\left(\left.\partial_{t}\vec{v}\right|_{\xi,\widetilde{\vec{v}}}\ftilde\right)_{\xi,t}\right\rangle _{\widetilde{\vec{v}}}=\left\langle 1,\partial_{t}\left(\frac{v^{2}}{2}J_{\xi}\widetilde{f}\right)_{\xi,\widetilde{\vec{v}}}\right\rangle _{\widetilde{\vec{v}}}.\label{eq:temporal_continuum_energy_constraint}
\end{equation}

In the discrete, the symmetries given by Eqs. (\ref{eq:temporal_continuum_momentum_constraint})
and (\ref{eq:temporal_continuum_energy_constraint}) are not satisfied
automatically due to truncation error, ${\cal O}\left(\Delta_{v}^{\beta},\Delta_{t}^{\zeta}\right)$,
where $\beta$ and $\zeta$ denote the order of discretization in
the velocity space and time, respectively. We ensure them by multiplying
the inertial term by a velocity-space dependent function, $\gamma_{t}\left(\vec{v}\right)$,
such that, at a time-step $p$, 

\begin{eqnarray}
\left\langle v_{||,i}^{(p)},\delta_{\widetilde{\vec{v}}}\cdot\left(\gamma_{t,i}^{(p+1)}\underbrace{J_{\xi,i}^{(p+1)}\left[v_{i}^{*,(p)}\right]^{-1}\delta_{t}\vec{v}_{i}^{(p+1)}}_{\textcircled a}\widetilde{f}_{i}^{(p+1)}\right)\right\rangle _{\delta\vec{\widetilde{v}}}-\label{eq:temporal_semi_discrete_momentum_constraint}\\
\left[\left\langle v_{||,i}^{(p)},\delta_{t}\left(J_{\xi}\widetilde{f}\right)_{i}^{(p+1)}\right\rangle _{\delta\vec{\widetilde{v}}}-\left\langle 1,\delta_{t}\left(v_{||}J_{\xi}\widetilde{f}\right)_{i}^{(p+1)}\right\rangle _{\delta\vec{\widetilde{v}}}\right]=0
\end{eqnarray}
and
\begin{eqnarray}
\left\langle \frac{\left[v_{i}^{(p)}\right]^{2}}{2},\delta_{\widetilde{\vec{v}}}\cdot\left(\gamma_{t,i}^{(p+1)}J_{\xi,i}^{(p+1)}\left[v_{i}^{*,(p)}\right]^{-1}\delta_{t}\vec{v}_{i}^{(p+1)}\ftilde_{i}^{(p+1)}\right)\right\rangle _{\delta\widetilde{\vec{v}}}-\nonumber \\
\left[\left\langle \frac{\left[v_{i}^{(p)}\right]^{2}}{2},\delta_{t}\left(J_{\xi}\widetilde{f}\right)_{i}^{(p+1)}\right\rangle _{\delta\widetilde{\vec{v}}}-\left\langle 1,\delta_{t}\left(\frac{v^{2}}{2}J_{\xi}\widetilde{f}\right)_{i}^{(p+1)}\right\rangle _{\delta\widetilde{\vec{v}}}\right]=0\label{eq:temporal_semi_discrete_energy_constraint}
\end{eqnarray}
Here, $i$ is the spatial cell index (velocity-space indices are dropped
for brevity), $\delta_{\widetilde{\vec{v}}}\cdot$ is the discrete
velocity-space divergence operator, $\delta_{t}$ is the discrete
temporal derivative, $\gamma_{t}\left(\vec{\widetilde{v}}\right)=1+{\cal O}\left(\Delta_{v}^{\beta},\Delta_{t}^{\zeta}\right)$
is the discrete-nonlinear-constraint function \citep{Taitano_2018_vrfp_1d2v_implicit},
and
\[
\vec{v}_{i}^{(p)}=v_{i}^{*,(p)}\left[\widetilde{\vec{v}}+\widehat{u}_{||}^{*,(p)}\vec{e}_{||}\right],
\]
\[
\delta_{t}\vec{v}_{i}^{(p+1)}=\frac{c^{(p+1)}\vec{v}_{i}^{(p)}+c^{(p)}\vec{v}_{i}^{(p-1)}+c^{(p-1)}\vec{v}_{i}^{(p-2)}}{\Delta t^{(p)}},
\]
\[
\delta_{t}\left(J_{\xi}\widetilde{f}\right)_{i}^{(p+1)}=\frac{c^{(p+1)}J_{\xi,i}^{(p+1)}\widetilde{f}_{i}^{(p+1)}+c^{(p)}J_{\xi,i}^{(p)}\widetilde{f}_{i}^{(p)}+c^{(p-1)}J_{\xi,i}^{(p-1)}\widetilde{f}_{i}^{(p-1)}}{\Delta t^{(p)}},
\]
\[
\delta_{t}\left(v_{||}J_{\xi}\widetilde{f}\right)_{i}^{(p+1)}=\frac{c^{(p+1)}v_{||,i}^{(p)}J_{\xi,i}^{(p+1)}\widetilde{f}_{i}^{(p+1)}+c^{(p)}v_{||,i}^{(p-1)}J_{\xi,i}^{(p)}\widetilde{f}_{i}^{(p)}+c^{(p-1)}v_{||,i}^{(p-2)}J_{\xi,i}^{(p-1)}\widetilde{f}_{i}^{(p-1)}}{\Delta t^{(p)}},
\]
\[
\delta_{t}\left(\frac{v^{2}}{2}J_{\xi}\widetilde{f}\right)_{i}^{(p+1)}=\frac{c^{(p+1)}\left[v_{i}^{(p)}\right]^{2}\widetilde{f}_{i}^{(p+1)}+c^{(p)}\left[v_{i}^{(p-1)}\right]^{2}\widetilde{f}_{i}^{(p)}+c^{(p-1)}\left[v_{i}^{(p-2)}\right]^{2}\widetilde{f}_{i}^{(p-1)}}{2\Delta t^{(p)}}.
\]
We remind the readers that $\textcircled a$ in Eq. (\ref{eq:temporal_semi_discrete_momentum_constraint})
is the advection velocity shown in Eqs. (\ref{eq:temporal_inertial_coeff_parallel})
and (\ref{eq:temporal_inertial_coeff_perpendicular}). In order to
evaluate $\gamma_{t}$, we follow Ref. \citep{Taitano_2018_vrfp_1d2v_implicit}
and begin by assuming a functional representation:

\begin{equation}
\gamma_{t}\left(v_{||},v_{\perp}\right)=1+\sum_{l=0}^{P}C_{l}B_{l}\left(v_{||},v_{\perp}\right),\label{eq:temporal_discrete_nonlinear_constraint_function}
\end{equation}
where
\[
\sum_{l=0}^{P}C_{l}B_{l}\left(v_{||},v_{\perp}\right)=\sum_{r=0}^{P_{||}}\sum_{s=0}^{P_{\perp}}C_{rs}B_{||,r}\left(v_{||}\right)B_{\perp,s}\left(v_{\perp}\right),
\]
$B_{||,r}$ is a $r^{th}$ functional representation in the parallel
velocity component, $B_{\perp,s}$ is a similar quantity in the perpendicular
velocity component, and $C_{rs}$ is the coefficient corresponding
to the respective functions. In this study, we chose a Fourier representation
where: 
\[
B_{||,r}=\begin{cases}
1 & \textnormal{if}\;r=0\\
\textnormal{sin}\left[rk_{||}\left(v_{||}-u_{||}\right)\right] & \textnormal{if}\;\textnormal{mod}\left(r,2\right)=0\\
\textnormal{cos}\left[\left(r-1\right)k_{||}\left(v_{||}-u_{||}\right)\right] & \textnormal{if}\;\textnormal{mod}\left(r,2\right)=1
\end{cases},\;\;\;B_{\perp,s}=\begin{cases}
1 & \textnormal{if}\;s=0\\
\textnormal{sin}\left[sk_{\perp}v_{\perp}\right] & \textnormal{if}\;\textnormal{mod}\left(s,2\right)=0\\
\textnormal{cos}\left[\left(s-1\right)k_{\perp}v_{\perp}\right] & \textnormal{if}\;\textnormal{mod}\left(s,2\right)=1
\end{cases},
\]
and $k_{||}=2\pi/L_{||}$, $k_{\perp}=2\pi/L_{\perp}$ are the wave
vectors. We also choose $r=s=\left(0,1,2\right)$. The coefficients,
$C_{l}$, are obtained by minimizing their amplitude while satisfying
the discrete symmetry constraints as given by Eqs. (\ref{eq:temporal_semi_discrete_momentum_constraint})
and (\ref{eq:temporal_semi_discrete_energy_constraint}). This is
done by solving a constrained-minimization problem for the following
objective function:

\begin{equation}
{\cal F}\left(\vec{C},\bm{\lambda}\right)=\frac{1}{2}\sum_{l=0}^{P}C_{l}^{2}-\bm{\lambda}^{T}\cdot\vec{M}.\label{eq:cost_function_t}
\end{equation}
Here, $\vec{C}=\left[C_{1},C_{2},\cdots,C_{P}\right]^{T}$, $\bm{\lambda}$
is a vector of Lagrange multipliers and $\vec{M}$ is the vector of
vanishing constraints {[}Eqs. (\ref{eq:temporal_semi_discrete_momentum_constraint})
and (\ref{eq:temporal_semi_discrete_energy_constraint}){]}; and $\vec{C}$
is obtained from the linear system:

\begin{equation}
\left[\begin{array}{c}
\partial_{\bm{C}}{\cal F}\\
\partial\bm{\lambda}{\cal F}
\end{array}\right]=\vec{0}.\label{eq:minimization_problem}
\end{equation}
We note that similarly to previous studies employing discrete nonlinear
constraints \citep{taitano-jcp-15-vfp,Taitano_2015_jcp_cmec_va,Taitano_2015_rfp_0d2v_implicit,Taitano_2016_rfp_0d2v_implicit,Taitano_2018_vrfp_1d2v_implicit},
since $\gamma_{t}$ is an implicit function of the solution, for nonlinearly
implicit system - such as ours - the quality of discrete conservation
properties depends on the prescribed nonlinear convergence tolerance
of our solver (as demonstrated in Sec. \ref{subsec:numerical_results_1d2v_sin_relax}).

\subsubsection{Spatial variation of $v^{*}$ and $\widehat{u}_{||}^{*}$}

Similarly to the temporal variation, conservation symmetries for the
case of spatial variation of the metrics can be shown independently.
Consider only the spatial gradient terms in the Vlasov equation, Eq.
(\ref{eq:vfp_eqn_transformed}), to obtain the following expression:

\begin{equation}
\partial_{\xi}\left(v_{||,eff}\widetilde{f}\right)_{\vec{\widetilde{v}},t}-\frac{1}{v^{*}}\frac{\partial}{\partial\widetilde{\vec{v}}}\cdot\left(\left.\partial_{\xi}\vec{v}\right|_{\vec{\widetilde{v}},t}v_{||,eff}\ftilde\right)_{\xi,t}.\label{eq:continuum_spatial_inertial_vlasov}
\end{equation}
Here, $v_{||,eff}=v^{*}\left(\widetilde{v}_{||}+\widehat{u}_{||}^{*}\right)-\dot{x}$
and $\left.\partial_{\xi}\vec{v}\right|_{\vec{\widetilde{v}},t}=\partial_{\xi}\left[v^{*}\left(\vec{v}+\widehat{u}_{||}^{*}\vec{e}_{||}\right)\right]_{\widetilde{\vec{v}},t}$,
and the mass conservation theorem is revealed by taking the $mv^{0}$
moment, assuming a periodic boundary condition, and integrating in
$\xi$ to find
\begin{equation}
\int_{0}^{1}\left\langle 1,\partial_{\xi}\left(mv_{||,eff}\widetilde{f}\right)\right\rangle _{\widetilde{\vec{v}}}d\xi=0.
\end{equation}
Note that the inertial term is in a divergence form in the velocity
space, and therefore, its $mv^{0}$ moment trivially vanishes both
continuously and discretely.

For momentum conservation, we require,
\begin{equation}
\int_{0}^{1}d\xi\left\langle 1,\partial_{\xi}\left[mv_{||}v_{||,eff}\widetilde{f}\right]\right\rangle _{\widetilde{\vec{v}}}=0.\label{eq:spatial_momentum_constraint}
\end{equation}
This can be shown by multiplying Eq. (\ref{eq:continuum_spatial_inertial_vlasov})
by $mv_{||}$ and using the chain-rule to obtain 

\begin{equation}
\partial_{\xi}\left(mv_{||}v_{||,eff}\widetilde{f}\right)_{\vec{\widetilde{v}},t}-m\left\{ \underbrace{v_{||,eff}\widetilde{f}\left.\partial_{\xi}v_{||}\right|_{\vec{\widetilde{v}},t}}_{\textcircled a}+\underbrace{v_{||}\frac{\partial}{\partial\widetilde{\vec{v}}}\cdot\left(\left[v^{*}\right]^{-1}\left.\partial_{\xi}\vec{v}\right|_{\vec{\widetilde{v}},t}v_{||,eff}\ftilde\right)_{\xi,t}}_{\textcircled b}\right\} .\label{eq:spatial_momentum_conservation_symmetry}
\end{equation}
By taking the velocity-space moment, $\left\langle 1,\left(\cdot\right)\right\rangle _{\widetilde{\vec{v}}}$,
of the above expression, terms $\textcircled a$ and $\textcircled b$
yields:

\[
\left\langle 1,\textcircled a\right\rangle _{\widetilde{\vec{v}}}=\left\langle 1,v_{||,eff}\widetilde{f}\left.\partial_{\xi}v_{||}\right|_{\widetilde{v}_{||},t}\right\rangle _{\widetilde{\vec{v}}},
\]
\[
\left\langle 1,\textcircled b\right\rangle _{\widetilde{\vec{v}}}=\left\langle \frac{v_{||}}{v^{*}},\frac{\partial}{\partial\widetilde{\vec{v}}}\cdot\left(\left.\partial_{\xi}\vec{v}\right|_{\vec{\widetilde{v}},t}v_{||,eff}\ftilde\right)_{\xi,t}\right\rangle _{\widetilde{\vec{v}}}=-\left\langle 1,v_{||,eff}\ftilde\left.\partial_{\xi}v_{||}\right|_{\widetilde{v}_{||},t}\right\rangle _{\widetilde{\vec{v}}},
\]
canceling each other and leaving alone the conservative term, $\left\langle 1,\partial_{\xi}\left[mv_{||}v_{||,eff}\widetilde{f}\right]_{\vec{\widetilde{v}},t}\right\rangle _{\widetilde{\vec{v}}}$,
and trivially satisfying Eq. (\ref{eq:spatial_momentum_constraint}).
In other words we must ensure the following exact relationship in
the discrete:

\begin{equation}
\int_{0}^{1}d\xi\left\{ \left\langle v_{||},\partial_{\xi}\left(v_{||,eff}\widetilde{f}\right)_{\vec{\widetilde{v}},t}\right\rangle _{\widetilde{\vec{v}}}-\left\langle v_{||},\frac{\partial}{\partial\widetilde{\vec{v}}}\cdot\left(\left[v^{*}\right]^{-1}\left.\partial_{\xi}\vec{v}\right|_{\vec{\widetilde{v}},t}v_{||,eff}\ftilde\right)_{\xi,t}\right\rangle _{\widetilde{\vec{v}}}\right\} =0.\label{eq:spatial_momentum_conservation_symmetry_2}
\end{equation}

For energy conservation, we require,

\begin{equation}
\int_{0}^{1}d\xi\left\langle 1,\partial_{\xi}\left[m\frac{v^{2}}{2}v_{||,eff}\widetilde{f}\right]\right\rangle _{\widetilde{\vec{v}}}=0.\label{eq:spatial_energy_constraint}
\end{equation}
This can be shown by multiplying Eq. (\ref{eq:continuum_spatial_inertial_vlasov})
by $m\frac{v^{2}}{2}$ and using the chain-rule to obtain
\begin{equation}
\partial_{\xi}\left[\frac{mv^{2}}{2}v_{||,eff}\widetilde{f}\right]-m\left\{ \underbrace{v_{||,eff}\widetilde{f}\partial_{\xi}\frac{v^{2}}{2}}_{\textcircled a}+\underbrace{\frac{v^{2}}{2}\frac{\partial}{\partial\widetilde{\vec{v}}}\cdot\left(\left[v^{*}\right]^{-1}\left.\partial_{\xi}\vec{v}\right|_{\widetilde{\vec{v}},t}v_{||,eff}\ftilde\right)}_{\textcircled b}\right\} .\label{eq:spatial_energy_conservation_symmetry}
\end{equation}
By taking the velocity-space moment, $\left\langle 1,\left(\cdot\right)\right\rangle _{\widetilde{\vec{v}}}$,
of the above expression, terms $\textcircled a$ and $\textcircled b$
yields:

\[
\left\langle 1,\textcircled a\right\rangle _{\widetilde{\vec{v}}}=\left\langle 1,v_{||,eff}\widetilde{f}\left.\partial_{\xi}\frac{v^{2}}{2}\right|_{\widetilde{\vec{v}},t}\right\rangle _{\widetilde{\vec{v}}},
\]
\[
\left\langle 1,\textcircled b\right\rangle _{\widetilde{\vec{v}}}=\left\langle \frac{v^{2}}{2v^{*}},\frac{\partial}{\partial\widetilde{\vec{v}}}\cdot\left(\left.\partial_{\xi}\vec{v}\right|_{\vec{\widetilde{v}},t}v_{||,eff}\ftilde\right)_{\xi,t}\right\rangle _{\widetilde{\vec{v}}}=-\left\langle \vec{v},v_{||,eff}\ftilde\left.\partial_{\xi}\vec{v}\right|_{\widetilde{\vec{v}},t}\right\rangle _{\widetilde{\vec{v}}}=-\left\langle 1,v_{||,eff}\ftilde\left.\partial_{\xi}\frac{v^{2}}{2}\right|_{\widetilde{\vec{v}},t}\right\rangle _{\widetilde{\vec{v}}},
\]
canceling each other and leaving alone the conservative term, $\left\langle 1,\partial_{\xi}\left[m\frac{v^{2}}{2}v_{||,eff}\widetilde{f}\right]_{\vec{\widetilde{v}},t}\right\rangle _{\widetilde{\vec{v}}}$,
and trivially satisfying Eq. (\ref{eq:spatial_energy_constraint}).
In other words we must ensure the following exact relationship in
the discrete:

\begin{equation}
\int_{0}^{1}d\xi\left\{ \left\langle \frac{v^{2}}{2},\partial_{\xi}\left(v_{||,eff}\widetilde{f}\right)_{\widetilde{\vec{v}},t}\right\rangle _{\widetilde{\vec{v}}}-\left\langle \frac{v^{2}}{2},\frac{1}{v^{*}}\frac{\partial}{\partial\widetilde{\vec{v}}}\cdot\left(\left.\partial_{\xi}\vec{v}\right|_{\widetilde{v},t}v_{||,eff}\ftilde\right)_{\widetilde{\vec{v}},t}\right\rangle _{\widetilde{\vec{v}}}\right\} =0.\label{eq:spatial_energy_conservation_symmetry_2}
\end{equation}

In the discrete, similarly to the temporal symmetries, Eqs. (\ref{eq:spatial_momentum_conservation_symmetry_2})
and (\ref{eq:spatial_energy_conservation_symmetry_2}) are not satisfied
automatically due to truncation error, ${\cal O}\left(\Delta_{v}^{\beta},\Delta_{x}^{\eta}\right)$,
where $\beta$ and $\eta$ denote the order of discretization in the
velocity space and configuration space, respectively. We ensure these
relationships by modifying the inertial term by a velocity-space dependent
function, $\gamma_{x}\left(\vec{v}\right)$, such that,

\begin{eqnarray}
\sum_{i=1}^{N_{\xi}}\Delta\xi\left\{ \left\langle v_{||,i}^{(p)},\frac{\left(v_{||,eff}\widetilde{f}\right)_{i+1/2}^{(p+1)}-\left(v_{||,eff}\widetilde{f}\right)_{i-1/2}^{(p+1)}}{\Delta\xi}\right\rangle _{\delta\widetilde{\vec{v}}}\right.\nonumber \\
\left.-\left\langle v_{||,i}^{(p)},\frac{1}{2}\delta_{\widetilde{\vec{v}}}\cdot\left(\gamma_{x,i+1/2}^{(p+1)}\underbrace{\frac{\vec{v}_{i+1}^{(p)}-\vec{v}_{i}^{(p)}}{v_{i}^{*,(p)}\Delta\xi}v_{||,eff,i+1/2}^{(p)}}_{\textcircled a}\ftilde_{i}^{(p+1)}+\gamma_{x,i-1/2}^{(p+1)}\underbrace{\frac{\vec{v}_{i}^{(p)}-\vec{v}_{i-1}^{(p)}}{v_{i}^{*,(p)}\Delta\xi}v_{||,eff,i-1/2}^{(p)}}_{\textcircled b}\ftilde_{i}^{(p+1)}\right)\right\rangle _{\delta\widetilde{\vec{v}}}\right\} =0\label{eq:spatial_semi_discrete_momentum_constraint}
\end{eqnarray}
and
\begin{eqnarray}
\sum_{i=1}^{N_{\xi}}\Delta\xi\left\{ \left\langle \frac{\left[v_{i}^{(p)}\right]^{2}}{2},\frac{\left(v_{||,eff}\widetilde{f}\right)_{i+1/2}^{(p+1)}-\left(v_{||,eff}\widetilde{f}\right)_{i-1/2}^{(p+1)}}{\Delta\xi}\right\rangle _{\delta\widetilde{\vec{v}}}\right.\nonumber \\
\left.-\left\langle \frac{\left[v_{i}^{(p)}\right]^{2}}{2},\frac{1}{2}\delta_{\widetilde{\vec{v}}}\cdot\left(\gamma_{x,i+1/2}^{(p+1)}\frac{\vec{v}_{i+1}^{(p)}-\vec{v}_{i}^{(p)}}{v_{i}^{*,(p)}\Delta\xi}v_{||,eff,i+1/2}^{(p)}\ftilde_{i}^{(p+1)}+\gamma_{x,i-1/2}^{(p+1)}\frac{\vec{v}_{i}^{(p)}-\vec{v}_{i-1}^{(p)}}{v_{i}^{*,(p)}\Delta\xi}v_{||,eff,i-1/2}^{(p)}\ftilde_{i}^{(p+1)}\right)\right\rangle _{\delta\widetilde{\vec{v}}}\right\} =0.\label{eq:spatial_semi_discrete_energy_constraint}
\end{eqnarray}
Here, $\left(v_{||,eff}\widetilde{f}\right)_{i+1/2}^{(p+1)}=F_{x,i+1/2}^{(p+1)}+F_{\dot{x},i+1/2}^{(p+1)}$
and $\vec{v}_{i}^{(p)}=v_{i}^{*,(p)}\left[\widetilde{\vec{v}}+\widehat{u}_{||,i}^{*,(p)}\vec{e}_{||}\right]$.
We point out that $\textcircled a$ and $\textcircled b$ in Eq. (\ref{eq:spatial_semi_discrete_momentum_constraint})
are the advection velocities given in Eqs. (\ref{eq:discrete_spatial_inertial_flux_parallel_coefficient})
and (\ref{eq:discrete_spatial_inertial_flux_perpendicular_coefficient});
and $\gamma_{x}\left(\vec{\widetilde{v}}\right)=1+{\cal O}\left(\Delta_{v}^{\beta},\Delta_{x}^{\eta}\right)$
is the spatial discrete-nonlinear-constraint function where the functional
form is chosen to be similarly to $\gamma_{t}$ {[}Eq. (\ref{eq:temporal_discrete_nonlinear_constraint_function}){]}
in this study. Performing a discrete integration by parts (i.e., telescoping
of the summation) on Eqs. (\ref{eq:spatial_semi_discrete_momentum_constraint})
and (\ref{eq:spatial_semi_discrete_energy_constraint}), we obtain
the following constraints that relates $\gamma_{x,i+1/2}$, the discrete
configuration-space flux, and the velocity-space inertial terms:
\begin{eqnarray}
\left\langle v_{||,i}^{(p)}-v_{||,i+1}^{(p)},\left(v_{||,eff}\widetilde{f}\right)_{i+1/2}^{(p+1)}\right\rangle _{\delta\widetilde{\vec{v}}}\nonumber \\
-\frac{1}{2}\left[\left\langle v_{||,i}^{(p)},\delta_{\widetilde{\vec{v}}}\cdot\left(\gamma_{x,i+1/2}^{(p+1)}\left[v_{i}^{*,(p)}\right]^{-1}\left[\vec{v}_{i+1}^{(p)}-\vec{v}_{i}^{(p)}\right]v_{||,eff,i+1/2}^{(p)}\ftilde_{i}^{(p+1)}\right)\right\rangle _{\delta\widetilde{\vec{v}}}\right.\nonumber \\
\left.+\left\langle v_{||,i+1}^{(p)},\delta_{\widetilde{\vec{v}}}\cdot\left(\gamma_{x,i+1/2}^{(p+1)}\left[v_{i+1}^{*,(p)}\right]^{-1}\left[\vec{v}_{i+1}^{(p)}-\vec{v}_{i}^{(p)}\right]v_{||,eff,i+1/2}^{(p)}\ftilde_{i+1}^{(p+1)}\right)\right\rangle _{\delta\widetilde{\vec{v}}}\right]=0\label{eq:semi_discrete_spatial_momentum_constraint_3}
\end{eqnarray}
and
\begin{eqnarray}
\left\langle \frac{\left[v_{i}^{(p)}\right]^{2}}{2}-\frac{\left[v_{i+1}^{(p)}\right]^{2}}{2},\left(v_{||,eff}\widetilde{f}\right)_{i+1/2}^{(p+1)}\right\rangle _{\delta\widetilde{\vec{v}}}\nonumber \\
-\frac{1}{2}\left[\left\langle \frac{\left[v_{i}^{(p)}\right]^{2}}{2},\delta_{\widetilde{\vec{v}}}\cdot\left(\gamma_{x,i+1/2}\left[v_{i}^{*,(p)}\right]^{-1}\left[\vec{v}_{i+1}^{(p)}-\vec{v}_{i}^{(p)}\right]v_{||,eff,i+1/2}^{(p)}\ftilde_{i}^{(p+1)}\right)\right\rangle _{\delta\widetilde{\vec{v}}}\right.\nonumber \\
\left.+\left\langle \frac{\left[v_{i+1}^{(p)}\right]^{2}}{2},\delta_{\widetilde{\vec{v}}}\cdot\left(\gamma_{x,i+1/2}\left[v_{i+1}^{*,(p)}\right]^{-1}\left[\vec{v}_{i+1}^{(p)}-\vec{v}_{i}^{(p)}\right]v_{||,eff,i+1/2}^{(p)}\ftilde_{i+1}^{(p+1)}\right)\right\rangle _{\delta\widetilde{\vec{v}}}\right]=0.\label{eq:semi_discrete_spatial_energy_constraint_3}
\end{eqnarray}
The vector of coefficients, $\vec{C}$, for $\gamma_{x,i+1/2}$ is
evaluated by solving a constrained minimization problem as in Eq.
(\ref{eq:minimization_problem}) with the vector of vanishing constraints,
$\vec{M}$, being Eqs. (\ref{eq:semi_discrete_spatial_momentum_constraint_3})
and (\ref{eq:semi_discrete_spatial_energy_constraint_3}). We end
the section by noting that at Dirichlet boundaries, we set $\gamma_{x,i+1/2}=1$
as the boundary condition violates the continuum conservation principle. 

\subsection{Discretization of the MMPDE and grid velocity (a null-space preserving
scheme)\label{subsec:grid_velocity_discretization}}

To solve the MMPDE, Eq. (\ref{eq:mmpde_equation}), for the new time
configuration-space coordinate, we discretize it using BDF2 in time
and standard finite differences in space:
\begin{equation}
\frac{c^{(p+1)}x_{i+1/2}^{(p+1)}+c^{(p)}x_{i+1/2}^{(p)}+c^{(p-1)}x_{i+1/2}^{(p-1)}}{\Delta t^{(p)}}-\tau_{x}^{-1}\frac{\omega_{x,i+1}^{(p)}\left(x_{i+3/2}^{(p+1)}-x_{i+1/2}^{(p+1)}\right)-\omega_{x,i}^{(p)}\left(x_{i+1/2}^{(p+1)}-x_{i-1/2}^{(p+1)}\right)}{\Delta\xi^{2}}=0,\label{eq:discrete_mmpde}
\end{equation}
where

\[
\omega_{x,i}^{(p)}=\omega_{x,i}^{(p)}\left({\cal M}_{j}^{(p)}\right)\;\textnormal{for}\;j\in\left[1,N_{\xi}\right]\;\textnormal{and}\;{\cal M}=\left(n_{\alpha}^{(p)},u_{||,\alpha}^{(p)},v_{th,\alpha}^{(p)},T_{e}^{(p)}\right)\;\textnormal{for}\;\text{\ensuremath{\alpha\in[1,N_{sp}].}}
\]
We note, that to avoid over-constraining the resulting nonlinear system
of equations arising from the discretized VFP and fluid electron system,
the monitor function is evaluated using quantities from the previous
time step.

As in most moving grid scheme, spatial-mesh smoothing is required
for both stability and accuracy of the solution. The smoothing ensures
that a relative variation of the grid size is smooth. In this study,
we choose a Winslow smoothing \citep{winslow_jcp_winslow_smoothing}
on the monitor function:

\begin{equation}
\left[1-\lambda_{\omega}\partial_{\xi\xi}^{2}\right]\omega_{x}=\omega_{x}^{*}.\label{eq:winslow_smoothing}
\end{equation}
Here, $\omega_{x}$ is the smoothed monitor function, $\omega_{x}^{*}$
is the original (pre-smoothed) quantity computed as discussed in Sec.
\ref{sec:MMPDE}, and $\lambda_{\omega}$ is an empirically chosen
($5\times10^{-3}$ in this study) positive scalar (i.e., the smoothing
coefficient).

The grid velocity in configuration space is computed to ensure the
null-space of the Vlasov operator in the limit of a homogeneous plasma.
Consider the following linear-advection equation,
\[
\frac{\partial\phi}{\partial t}+v\frac{\partial\phi}{\partial x}=0,
\]
for a scalar variable, $\phi=\phi\left(x,t\right)$, defined on $x\in[0,1]$
and $t\in\left[0,\infty\right)$, where $v$ is a constant-advection
velocity. Here, the null-space of the operator is comprised of all
constant $\phi$ in $x$. We transform the equation via a strategy
similar to that described in Sec. \ref{sec:coordinate_transformation}
as:
\[
\frac{\partial}{\partial t}\left(J_{\xi}\phi\right)_{\xi}+\frac{\partial}{\partial\xi}\left[\left(v-\dot{x}\right)\phi\right]_{t}=0.
\]
For constant $\phi$, we find $\partial_{t}J_{\xi}=\partial_{\xi}\dot{x}$,
which is trivially satisfied. Discretizing using a BDF2 in time and
finite volume in space, we obtain:
\[
\frac{c^{(p+1)}J_{\xi,i}^{(p+1)}\phi_{i}^{(p+1)}+c^{(p)}J_{\xi,i}^{(p)}\phi_{i}^{(p)}+c^{(p-1)}J_{\xi,i}^{(p-1)}\phi_{i}^{(p-1)}}{\Delta t^{(p)}}-\frac{\dot{x}_{i+1/2}^{(p+1)}\phi_{i+1/2}^{(p+1)}-\dot{x}_{i-1/2}^{(p+1)}\phi_{i-1/2}^{(p+1)}}{\Delta\xi}=0.
\]
With $J_{\xi,i}=\frac{x_{i+1/2}-x_{i-1/2}}{\Delta\xi}$ and constant
$\phi$, we obtain
\[
\frac{c^{(p+1)}\left(x_{i+1/2}^{(p+1)}-x_{i-1/2}^{(p+1)}\right)+c^{(p)}\left(x_{i+1/2}^{(p)}-x_{i-1/2}^{(p)}\right)+c^{(p-1)}\left(x_{i+1/2}^{(p-1)}-x_{i-1/2}^{(p-1)}\right)}{\Delta t^{(p)}}-\left(\dot{x}_{i+1/2}^{(p+1)}-\dot{x}_{i-1/2}^{(p+1)}\right)=0.
\]
By equating coefficients with the same spatial indices, one finds
that the grid velocity must be defined as:
\begin{equation}
\dot{x}_{i+1/2}^{(p+1)}=\frac{c^{(p+1)}x_{i+1/2}^{(p+1)}+c^{(p)}x_{i+1/2}^{(p)}+c^{(p-1)}x_{i+1/2}^{(p-1)}}{\Delta t^{(p)}},\label{eq:grid_velocity}
\end{equation}
which is the same temporal scheme used elsewhere.

\subsection{Evolution strategy for the velocity-space grid: spatial-temporal
adaptivity of $v^{*}$ and $\widetilde{u}_{||}^{*}$\label{subsec:Evolution-equation-of-vstar}}

\begin{figure}
\begin{centering}
\includegraphics[height=5cm]{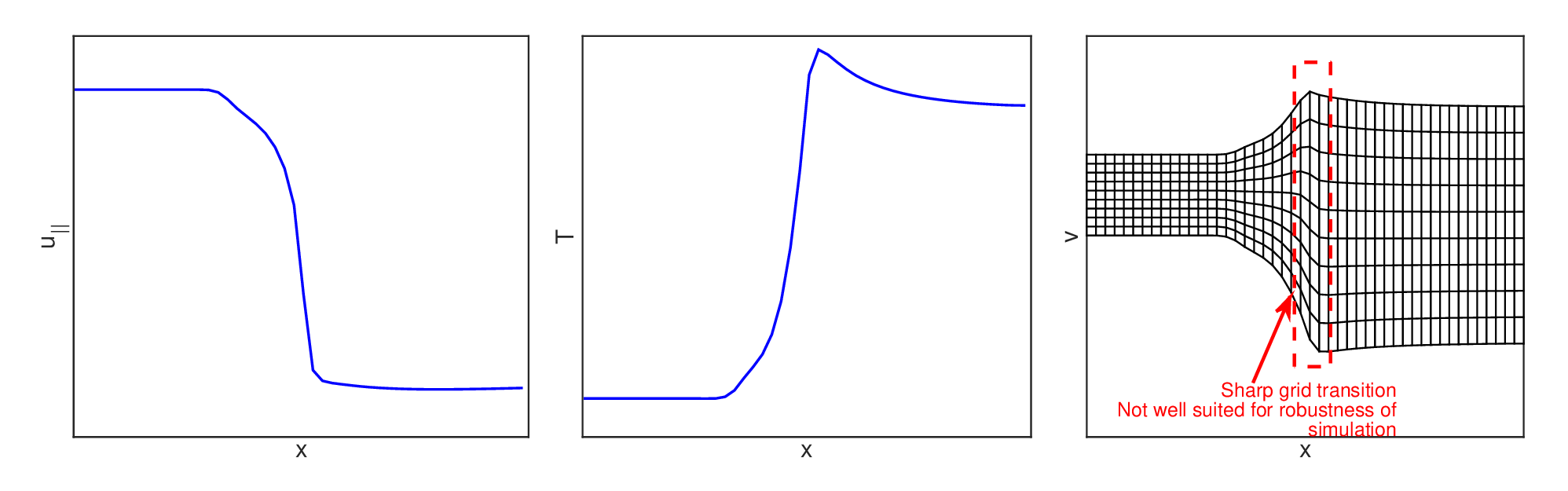}
\par\end{centering}
\caption{Illustration of plasma-shock flow velocity (left), temperature (center),
and the corresponding grid quality (right) near the shock front.\label{fig:illustration_1d_plasma_shock_T_and_mesh}}
\end{figure}

Similarly to the configuration space, a smooth variation of the  velocity-space
grid is a key for the robustness of the adaptivity strategy. We ensure
high-quality grids by employing a smoothing strategy for the temporal
and spatial variation of the normalization speed, $v^{*}$, and shift
velocity, $\widehat{u}_{||}^{*}$, used to transform the Vlasov-Fokker-Planck
equation. In Ref. \citep{Taitano_2018_vrfp_1d2v_implicit}, the Vlasov-Fokker-Planck
equation for each ion species was normalized to a quantity, which
is a function of its $v_{th}$, and was spatially smoothed using several
passes of binomial filtering. The spatial smoothing was necessary
to stabilize the grid-adaptivity scheme against numerical instabilities
where sharp variation in $v_{th}$ (e.g., plasma shocks) are present.
Consider a scenario where a planar plasma shock propagates through
a medium; refer to Fig. \ref{fig:illustration_1d_plasma_shock_T_and_mesh}.
For strong shocks, sharp temperature and drift velocity (in terms
of $v_{th}$) variations exist near the shock front. These large variations
in $v_{th}$ and $\widehat{u}_{||}$ will cause the velocity space
grid to be expanded/shifted rapidly both in space and time, resulting
in a numerical \emph{brittleness}. In this study, we address these
issues by combining: 1) an empirical temporal limiter, and 2) a spatial
smoothing operation, similarly to how we dealt with the monitor function.
We note, that neither of these strategies result in a loss of numerical
accuracy in principle, as the transformed equations are correct for
an arbitrary $v^{*}$ and $\widehat{u}_{||}^{*}$. We elaborate on
these strategies next.

In order to limit the velocity grid expansion/contraction and shifting
rate in time, we limit the change of updates of $v^{*}$ and $u_{||}^{*}=v^{*}\widehat{u}_{||}^{*}$
by $5\%$ from between time step, i.e.:

\begin{equation}
v_{\alpha}^{*,(p+1)}=\begin{cases}
v_{\alpha}^{*,(p)}+\Delta t^{(p)}\dot{v}_{\alpha}^{*,(p+1)} & \textnormal{if}\;\Delta t^{(p)}\frac{\left|\dot{v}_{\alpha}^{*,(p+1)}\right|}{v_{\alpha}^{*,(p)}}\le0.05\\
v_{\alpha}^{*,(p)}\left[1+0.05\Delta t\textnormal{sign}\left(\dot{v}_{\alpha}^{*,(p+1)}\right)\right] & \textnormal{otherwise}
\end{cases},\label{eq:vstar_def}
\end{equation}
and
\begin{equation}
u_{||,\alpha}^{*,(p+1)}=\begin{cases}
u_{||,\alpha}^{*,(p)}+\Delta t^{(p)}\dot{u}_{||,\alpha}^{*,(p+1)} & \textnormal{if}\;\Delta t^{(p)}\left|\frac{\dot{u}_{||,\alpha}^{*,(p+1)}}{v_{\alpha}^{*,(p)}}\right|\le0.05\\
u_{||,\alpha}^{*,(p)}\left[1+0.05\Delta t\textnormal{sign}\left(\dot{u}_{||,\alpha}^{*,(p+1)}\right)\right] & \textnormal{otherwise}
\end{cases},\label{eq:urstar_def}
\end{equation}
where

\[
\dot{v}_{\alpha}^{*,(p+1)}=\frac{\left[v_{\alpha}^{*,(p+1)}\right]^{\dagger}-v_{\alpha}^{*,(p)}}{\Delta t^{(p)}},\;\dot{u}_{||,\alpha}^{*,(p+1)}=\frac{\left[u_{||,\alpha}^{*,(p+1)}\right]^{\dagger}-u_{||,\alpha}^{*,(p)}}{\Delta t^{(p)}},
\]
and

\begin{eqnarray*}
\left[v_{\alpha}^{*,(p+1)}\right]^{\dagger}=\sqrt{\frac{2T_{\alpha}^{(p+1)}}{m_{\alpha}}},\;\left[u_{||,\alpha}^{*,(p+1)}\right]^{\dagger}=u_{||,\alpha}^{(p+1)}+\Delta w_{||,\alpha}^{(p+1)},\\
\Delta w_{||,\alpha}^{(p+1)}=\frac{\left\langle \left(v_{||}^{(p)}-u_{||,\alpha}^{(p+1)}\right)\left(\vec{v}^{(p)}-\vec{u}_{\alpha}^{(p+1)}\right)^{2},f_{\alpha}^{(p+1)}\right\rangle _{v}}{\left\langle \left(\vec{v}^{(p)}-\vec{u}_{\alpha}^{(p+1)}\right)^{2},f_{\alpha}^{(p+1)}\right\rangle _{v}}.
\end{eqnarray*}
Here $\Delta w_{||}$ is a measure of \emph{skewness }in the distribution
function and is included to better account for non-thermal structures.\emph{ }

\begin{figure}
\begin{centering}
\includegraphics[height=5.5cm]{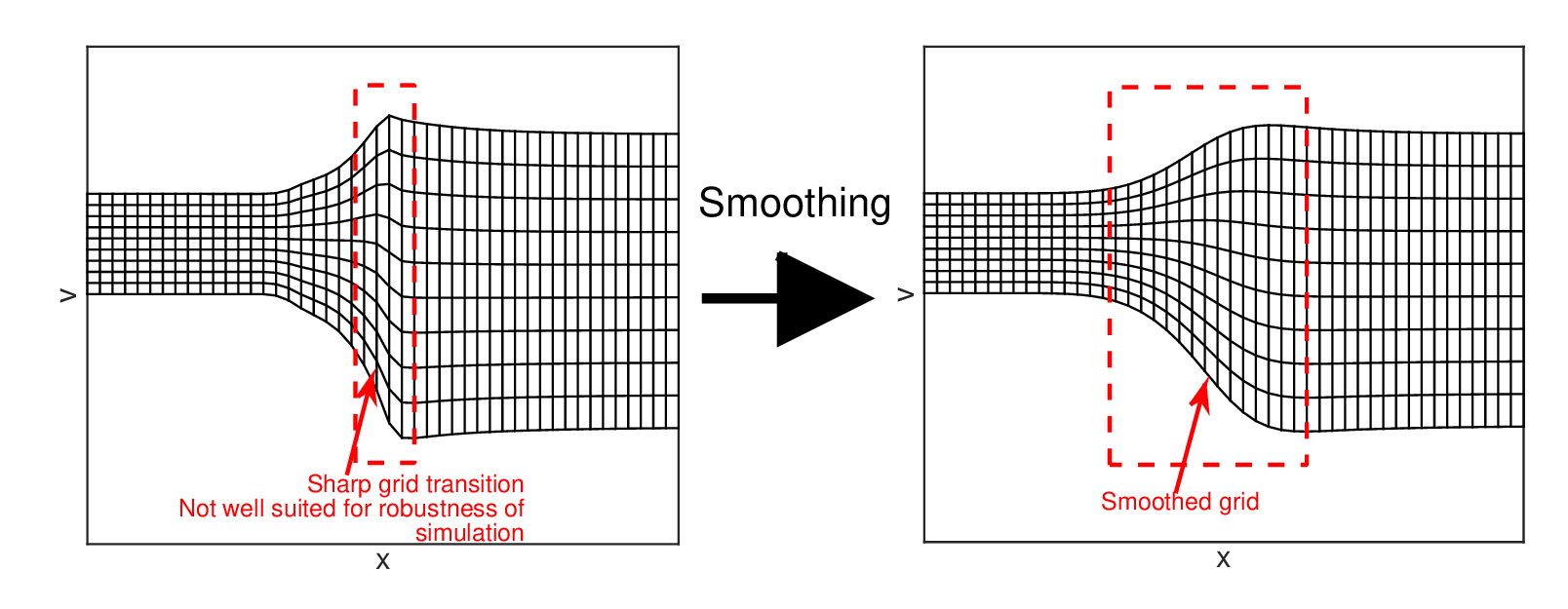}
\par\end{centering}
\caption{Illustration of unsmoothed (left) and smoothed (right) phase-space
grid in a plasma-shock problem. \label{fig:illustration_grid_smoothing}}
\end{figure}

To ensure that the profiles of $v^{*}$ and $u_{||}^{*}$ are smooth
\emph{in space}, we employ a Winslow smoother. We solve the following
elliptic equation for $\phi$,

\begin{equation}
\left[1-\lambda_{\phi}\partial_{\xi\xi}\right]\phi=\phi^{*}.
\end{equation}
Here, note that $\phi^{*}$ is the pre-smoothed quantity {[}$v^{*}$
and $u^{*}$ from Eqs. (\ref{eq:vstar_def}) and (\ref{eq:urstar_def}){]},
$\phi$ is the post-smoothed quantity, $\lambda_{\phi}=10^{-2}$ is
the empirically determined smoothing coefficient, and that $\widehat{u}_{||}^{*}=u_{||}^{*}/v^{*}$
is computed after smoothing $v^{*}$ and $u_{||}^{*}$. Refer to Fig.
\ref{fig:illustration_grid_smoothing} for an illustration of the
effects of post-grid-smoothing operation.

\section{Numerical Results\label{sec:numerical-results}}

In this section, we demonstrate the conservation, order of accuracy,
and computational savings properties of our numerical implementation,
with various examples of varying degrees of complexity. For all problems,
we normalize the mass, char\textcolor{black}{ge, temperature, density,
velocity, and time to the proton mass, $m$, proton charge, $e$,
reference temperature, $T^{0}$, density, $n^{0}$, characteristic
speed, $v^{0}=\sqrt{T^{0}/m}$, and time-scale, $\tau^{0}=\frac{3\sqrt{m}\left(T^{0}\right)^{3/2}}{4\sqrt{\pi}n^{0}\Lambda e^{4}}$,
respectively. A constant Coulomb logarithm, $\Lambda=10$, is used
everywhere unless otherwise stated. All normalized distribution functions
are initialized as Maxwellians, with prescribed moments in $n$, $u_{||}$,
and $T$ as:}

\textcolor{black}{
\begin{equation}
\widetilde{f}_{M}=\frac{n}{\pi^{3/2}}\left(\frac{v^{*}}{v_{th}}\right)^{3}\textnormal{exp}\left[-\frac{\left(v^{*}\widetilde{v}_{||}+u_{||}^{*}-u_{||}\right)^{2}+\left(v^{*}\widetilde{v}_{\perp}\right)^{2}}{v_{th}^{2}}\right].
\end{equation}
The initial profiles of the normalization speed, $v^{*}(\xi)$, and
shift velocity, $u_{||}^{*}\left(\xi\right)$, are found by applying
a Winslow smoothing (unless otherwise stated, $\lambda_{\phi}=10^{-2}$),
such that high wavenumber components of the initial temperature profile
(if present) are smoothed out to prevent large numerical errors stemming
from the computation of spatial gradients of $v^{*}$ and $u_{||}^{*}$
in the inertial term. We note that, in this study, we use a discrete
quadrature error accounting technique to ensure that discrete Maxwellian
moments agree exactly with the prescribed ones \citep{Taitano_JCP_2017_FP_equil_disc}.
Also, for post-processing purposes the cell-center location of configuration-space
is computed from a linear interpolation of cell-face location, $x_{i}=0.5(x_{i+1/2}+x_{i-1/2})$.}

\textcolor{black}{The nonlinearly-coupled discrete ion Vlasov-Fokker-Planck
and fluid electron equations are solved using an Anderson accelerated
fixed-point iterative solver \citep{anderson_aa_JACM_1965} with a
nonlinear elimination strategy for the Rosenbluth potentials (similar
to Ref. \citep{Taitano_2015_rfp_0d2v_implicit}) and similar preconditioning
strategies (multi-grid solver and operator-splitting) as discussed
in Refs. \citep{Taitano_2015_rfp_0d2v_implicit,Gasteiger_JOPP_2016_ADI_type_PC_for_SS_Vlasov}.
Finally, unless otherwise stated, we em}ploy a nonlinear convergence
tolerance of $\epsilon_{r}=10^{-3}$. 

\subsection{Particle-beam relaxation}

We test the basic capability of velocity-space grid adaptivity by
simulating a particle-beam relaxation in a background plasma. Specifically,
we consider a background Aluminum plasma ($m_{Al}=27$ and $q_{Al}=13$
) with a number density of $n_{Al,0}=1$, temperature of $T_{Al,0}=1$
($v_{th,Al}=0.272)$ and a drift velocity of $u_{||,Al,0}=0$; and
a Deuteron beam ($m_{D}=2$ and $q_{D}=1$) with an initial temperature
of $T_{D,0}=0.01$ ($v_{th,D}=0.1$), and a drift velocity of $u_{||,D,0}=30v_{th,Al}$.
We choose a velocity-space domain of $\widetilde{v}_{||}\times\widetilde{v}_{\perp}\in\left[-5,5\right]\times\left[0,5\right]$
and a mesh size of $N_{||}\times N_{\perp}=64\times32$.

We note that, for static grids, modeling of a particle-beam relaxation
is notoriously challenging due to the requirements of having to simultaneously
resolve the initially cold beam, and also being able to have a large
enough velocity-space domain to accommodate the intermediate broad
(hot) distribution function. If sufficient resolution is not provided
for the beam, important dynamic quantities, such as the velocity-space
parallel and perpendicular diffusion rates, cannot be captured accurately.

\begin{figure}[t]
\begin{centering}
\includegraphics[width=6cm]{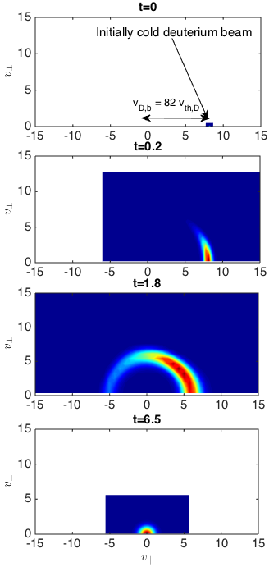}\includegraphics[width=6cm]{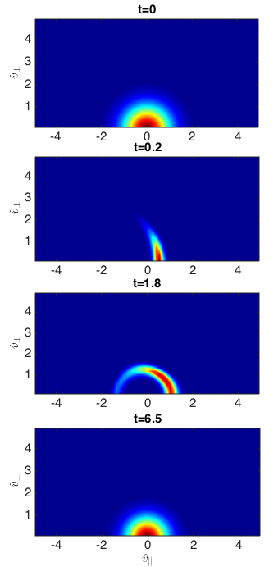}
\par\end{centering}
\caption{Particle-beam relaxation: Evolution of the beam distribution function
for various times in the physical (left column) and the transformed
coordinates (right column).}
\label{fig:0d2v_relaxation_processes}
\end{figure}

In Fig. \ref{fig:0d2v_relaxation_processes}, we show the evolution
of the beam distribution function for different snapshots in time.
As can be seen, the classic isotropization process takes place, followed
by the thermalization. It can also be seen that the grid dynamically
evolves with the temperature and drift velocities of the distribution
function, providing an optimal balance between grid resolution and
computational complexity at all times. 

\begin{figure}[t]
\begin{centering}
\includegraphics[height=6cm]{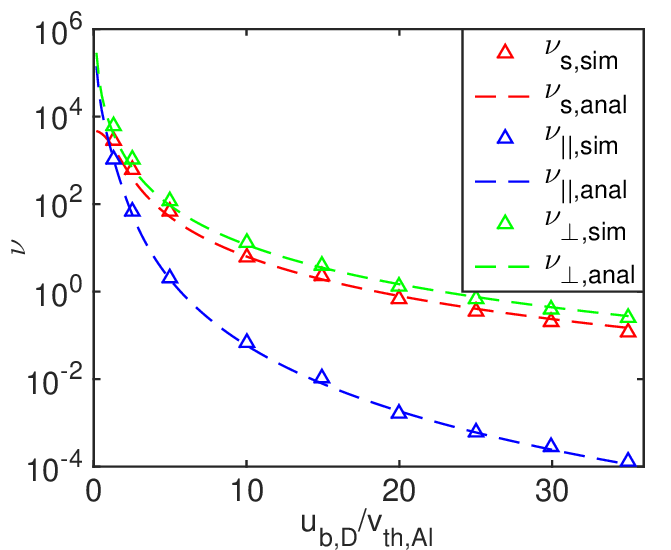}
\par\end{centering}
\caption{Particle beam relaxation: Comparison between numerical (triangles)
and theoretical (dashed lines) predictions for a Deuterium-beam slowing
down, parallel, and perpendicular diffusion rates on an Aluminum background. }
\label{fig:0d2v_relaxation_rates}
\end{figure}

Next, we demonstrate the accuracy of the new method by comparing the
beam slowing down, as well as the parallel and perpendicular diffusion
rates with theory predictions \citep{plasma-formulary},
\[
\nu_{s}^{D\backslash Al}=\left(1+m_{D}/m_{Al}\right)\psi\left(x^{D\backslash Al}\right)\nu_{0}^{D\backslash Al},
\]
\[
\nu_{\perp}^{D\backslash Al}=2\left[\left(1-1/2x^{D\backslash Al}\right)\psi\left(x^{D\backslash Al}\right)+\psi'\left(x^{D\backslash Al}\right)\right]\nu_{0}^{D\backslash Al},
\]
\[
\nu_{||}^{D\backslash Al}=\left[\psi\left(x^{D\backslash Al}\right)/x^{D\backslash Al}\right]\nu_{0}^{D\backslash Al},
\]
where
\[
\nu_{0}^{D\backslash Al}=\frac{4\pi e_{D}^{2}e_{Al}^{2}\Lambda_{DAl}n_{Al}}{m_{D}^{2}u_{||,D,0}^{3}},\;x^{D\backslash Al}=u_{||,D,0}^{2}/v_{th,Al,0}^{2},\;\psi=\frac{2}{\sqrt{\pi}}\int_{0}^{x}t^{1/2}e^{-t}dt,\;\psi'\left(x\right)=\frac{d\psi}{dx}.
\]
We time average the simulated quantities to increase numerical accuracy:

\[
\left\langle \overline{\nu_{s}^{D\backslash Al}}\right\rangle _{\tau}=\sum_{p=1}^{N_{t}}\left|\frac{u_{||,D}^{(p)}-u_{||,D}^{(p-1)}}{u_{||,D}^{(p)}}\right|/\sum_{p=1}^{N_{t}}\Delta t^{(p)},\;\textnormal{where}\;u_{||,D}^{(p)}=\frac{\left\langle v_{||}^{(p-1)},f_{D}^{(p)}\right\rangle _{v}}{\left\langle 1,f_{D}^{(p)}\right\rangle _{v}}
\]
\[
\left\langle \overline{\nu_{||}^{D\backslash Al}}\right\rangle _{\tau}=\sum_{p=1}^{N_{t}}\left|\frac{T_{||,D}^{(p)}-T_{||,D}^{(p-1)}}{\left[u_{||,D}^{(p)}\right]^{2}}\right|/\sum_{p=1}^{N_{t}}\Delta t^{(p)},\;\textnormal{where}\;T_{||,D}^{(p)}=\frac{m_{D}\left\langle \left(v_{||}^{(p-1)}-u_{||,D}^{(p)}\right)^{2},f_{D}^{(p)}\right\rangle _{v}}{\left\langle 1,f_{D}^{(p)}\right\rangle _{v}}.
\]

\[
\left\langle \overline{\nu_{\perp}^{D\backslash Al}}\right\rangle _{\tau}=\sum_{p=1}^{N_{t}}\left|\frac{T_{\perp,D}^{(p)}-T_{\perp,D}^{(p-1)}}{\left[u_{||,D}^{(p)}\right]^{2}}\right|/\sum_{p=1}^{N_{t}}\Delta t^{(p)},\;\textnormal{where}\;T_{\perp,D}^{(p)}=\frac{m_{D}\left\langle \left(v_{\perp}^{(p-1)}\right)^{2},f_{D}^{(p)}\right\rangle _{v}}{2\left\langle 1,f_{D}^{(p)}\right\rangle _{v}}.
\]
The results are shown in Fig. \ref{fig:0d2v_relaxation_rates} for
various initial Deuteron-beam velocities. Here, $N_{t}$ is the maximum
number of time steps chosen based on the averaging, which is performed
until 
\[
\left|T_{\perp,D}\left(t=t_{f}\right)-T_{\perp,D}\left(t=t_{0}\right)\right|/T_{\perp,D}\left(t=t_{0}\right)>0.01.
\]
Good agreement is seen across a wide range of beam velocities, demonstrating
the accuracy of the proposed grid-adaptivity scheme.

\subsection{Two-beam thermal and drift relaxation}

\begin{figure}[t]
\begin{centering}
\includegraphics[width=10cm]{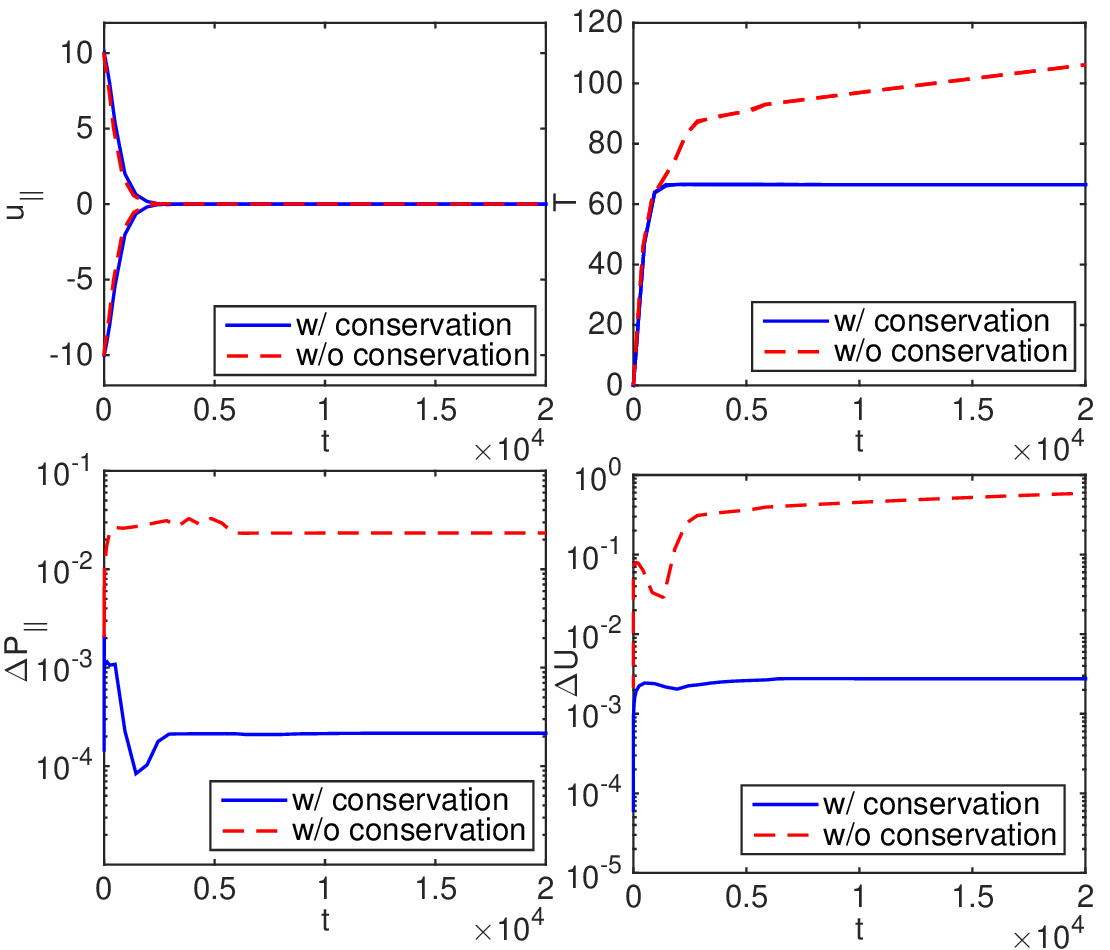}
\par\end{centering}
\caption{Two-beam thermal and drift relaxation: Dynamic relaxation of two species
temperature (top left) and drift velocity (top right) with (solid
line) and without (dashed line) discrete momentum (bottom left) and
energy (bottom right) conservation imposed. \label{fig:0d2v_two_species_thermal_drift_relaxation}}
\end{figure}

Next, we simulate relaxation of two counter-streaming Maxwellians
with disparate velocities. The purpose of this simulation is to demonstrate
the importance of satisfying the discrete temporal-conservation symmetries
arising from the velocity-space grid adaptivity {[}e.g., Eqs. (\ref{eq:temporal_semi_discrete_momentum_constraint})
and (\ref{eq:temporal_semi_discrete_energy_constraint}){]}. We compare
the solutions obtained with and without ensuring the symmetries.  We
consider two Deuteron beams ($q=1$ and $m=2$) with equal number
densities, $n=1$, and temperatures, $T=0.01$ ($v_{th}=0.1$), but
with disparate initial drift velocities, $u_{||,1}=-100v_{th}$ and
$u_{||,2}=100v_{th}$. We choose a velocity-space domain of $\widetilde{v}_{||}\times\widetilde{v}_{\perp}\in\left[-5,5\right]\times\left[0,5\right]$
and a mesh size of $N_{||}\times N_{\perp}=32\times16$. Through collisions,
the two beams will equilibrate to a common drift velocity of $u_{||}=0$
and a temperature of $T=66\frac{2}{3}$; refer to Fig. \ref{fig:0d2v_two_species_thermal_drift_relaxation}.
Here, 
\[
\Delta P_{||}=\left|\frac{\sum_{i}^{N_{\xi}}\sum_{\alpha}^{N_{sp}}J_{\xi,i}^{(p=0)}P_{||,\alpha,i}^{(p=0)}-\sum_{i}^{N_{\xi}}\sum_{\alpha}^{N_{sp}}J_{\xi,i}P_{||,\alpha,i}}{\sum_{i}^{N_{\xi}}\sum_{\alpha}^{N_{sp}}J_{\xi,i}^{(p=0)}P_{||,\alpha,i}^{(p=0)}}\right|
\]
is the relative variation of the total momentum of the system and
\[
\Delta U=\left|\frac{\sum_{i}^{N_{\xi}}\sum_{\alpha}^{N_{sp}}J_{\xi,i}^{(p=0)}U_{\alpha,i}^{(p=0)}-\sum_{i}^{N_{\xi}}\sum_{\alpha}^{N_{sp}}J_{\xi,i}U_{\alpha,i}}{\sum_{i}^{N_{\xi}}\sum_{\alpha}^{N_{sp}}J_{\xi,i}^{(p=0)}U_{\alpha,i}^{(p=0)}}\right|
\]
is the relative variation of the total kinetic energy of the system;
$P_{||,\alpha,i}=m_{\alpha}\left\langle v_{||,i},\widetilde{f}_{\alpha,i}\right\rangle _{\delta\widetilde{\vec{v}}}$
and $U_{\alpha,i}=m_{\alpha}\left\langle \frac{v_{,i}^{2}}{2},\widetilde{f}_{\alpha,i}\right\rangle _{\delta\widetilde{\vec{v}}}$.
Also, $i$ is the grid index in the configuration space, and the superscript
$p=0$ denotes the initial-time quantities. As can be seen, without
the temporal-Vlasov discrete-conservation symmetries, the system numerically
heats indefinitely. Thus, a moving-grid strategy is by itself, not
sufficient to reduce the computational complexity, as stable asymptotic
solutions cannot be achieved in general without a discrete conservation
principle. 

\subsection{1D2V sinusoidal perturbation relaxation with prescribed grid motion\label{subsec:numerical_results_1d2v_sin_relax}}

\begin{figure}[t]
\begin{centering}
\includegraphics[width=10cm]{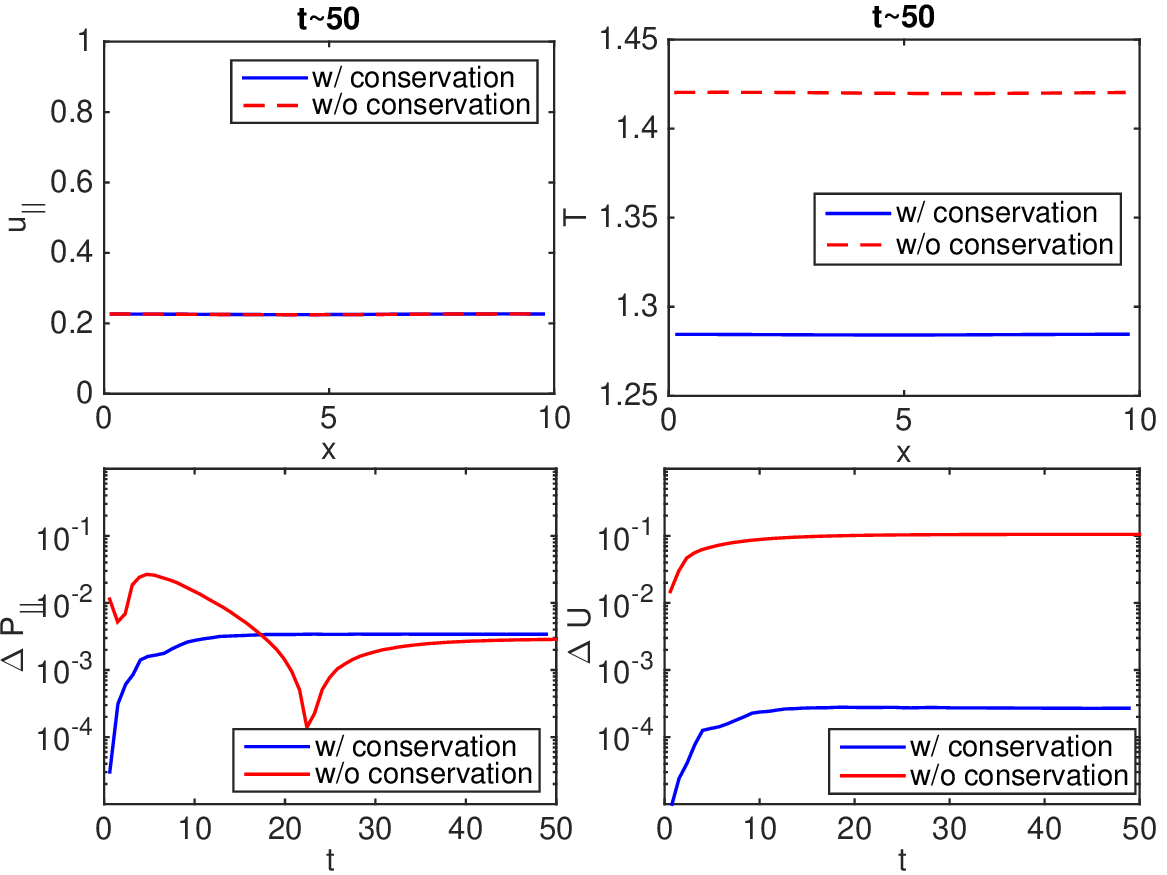}
\par\end{centering}
\caption{1D2V Sinusoidal perturbation relaxation: Dynamic relaxation of initial
drift velocity (top left) and temperature (top right) with (solid)
and without (dashed) the Vlasov discrete conservation symmetries,
and the associated momentum (bottom left) and energy (bottom right)
conservation error as a function of time. \label{fig:1D2V_Sinusoidal_wave_relaxation}}
\end{figure}

\begin{figure}[t]
\begin{centering}
\includegraphics[height=5cm]{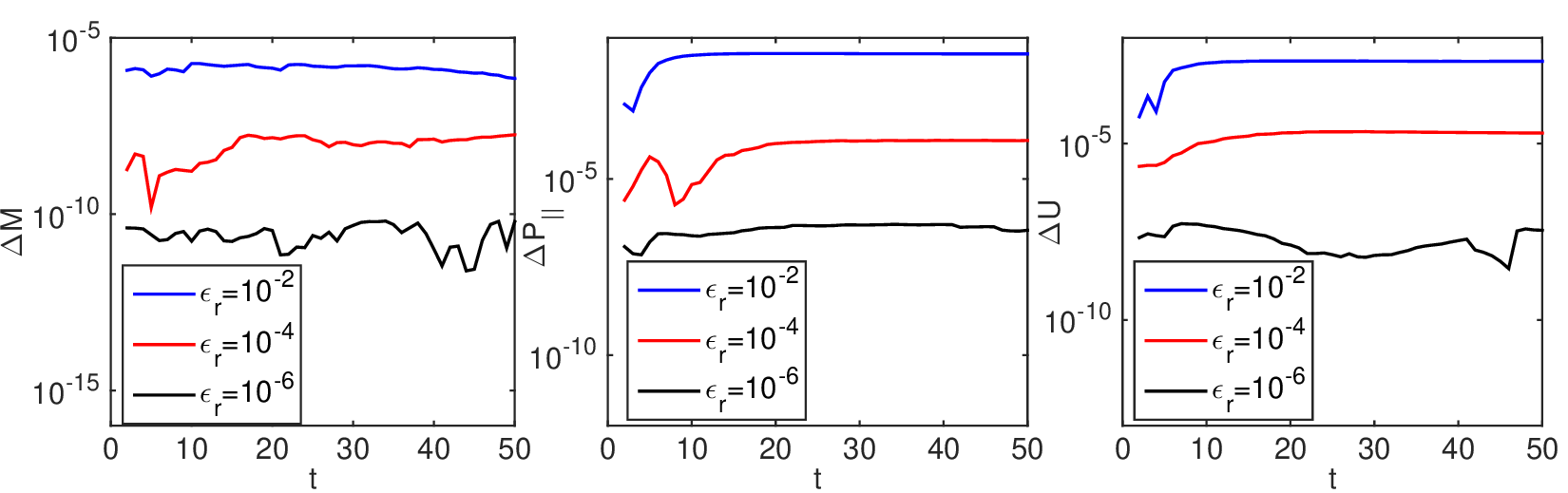}
\par\end{centering}
\caption{1D2V Sinusoidal perturbation relaxation: Conservation of mass, momentum,
and energy for various nonlinear convergence tolerances.\label{fig:1d2v_sin_wave_conservation_comparison}}
\end{figure}

We simulate a relaxation of a sinusoidal perturbation with a prescribed
grid motion. The purpose of this simulation is to demonstrate the
importance of satisfying the discrete conservation symmetries arising
from grid motion in configuration space. Consider a proton-electron
plasma ($q_{p}=1$, $q_{e}=-1$, $m_{p}=1$, and $m_{e}=1/1836$)
in a periodic domain with initial density, drift velocity, and temperature
profiles given by:

\[
n_{0}=1+0.5\sin\left(k_{x}x\right),
\]
\[
u_{||,0}=0.9\sin\left(k_{x}x\right),
\]
\[
T_{0}=1+0.9\sin\left(k_{x}x\right),
\]
where, $x\in[0,L_{x}]$, $L_{x}=10$ is the configuration-space domain
size, $k_{x}=2\pi/L_{x}$, and a time-dependent grid given by:

\[
x\left(\xi_{i},t\right)=x_{0,i}+0.45\Delta x_{0}\sin\left(\omega_{grid}t\right)\cos\left(k_{x}x_{0,i}\right).
\]
Here, $\Delta x_{0}=L_{x}/N_{\xi}$ is the average grid size, $N_{\xi}=24$,
$x_{0}=\xi L_{x}$ is the initial grid, $\xi\in\left[0,1\right]$
is the logical grid coordinate, and $\omega_{grid}=4\pi/5$ is the
grid oscillation frequency. In the velocity space, we employ the domain
$\widetilde{v}_{||}\times\widetilde{v}_{\perp}\in\left[-7,7\right]\times\left[0,7\right]$,
and the grid $N_{||}\times N_{\perp}=32\times16$. In Fig. \ref{fig:1D2V_Sinusoidal_wave_relaxation},
we compare the drift velocity and temperature of the protons at the
equilibrium time ($t\sim50$), with and without enforcing the Vlasov
temporal and spatial discrete conservation symmetries {[}Eqs. (\ref{eq:temporal_semi_discrete_momentum_constraint}),(\ref{eq:temporal_semi_discrete_energy_constraint}),(\ref{eq:semi_discrete_spatial_momentum_constraint_3}),
and (\ref{eq:semi_discrete_spatial_energy_constraint_3}){]}. 
As can be seen, the lack of discrete conservation manifests as numerical
heating, underlying the importance of satisfying the respective theorems.
In Fig. \ref{fig:1d2v_sin_wave_conservation_comparison}, we show
the quality of conservation with varying nonlinear-convergence tolerance;
and demonstrate that the quality improves with the tighter tolerance.

\begin{figure}[t]
\begin{centering}
\includegraphics[width=5cm]{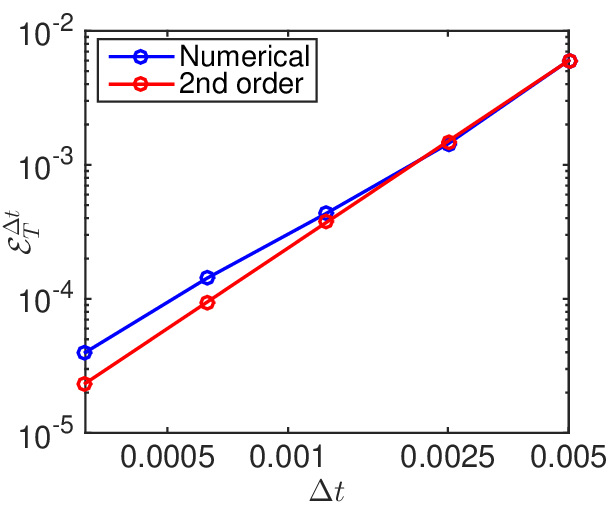}\includegraphics[width=5cm]{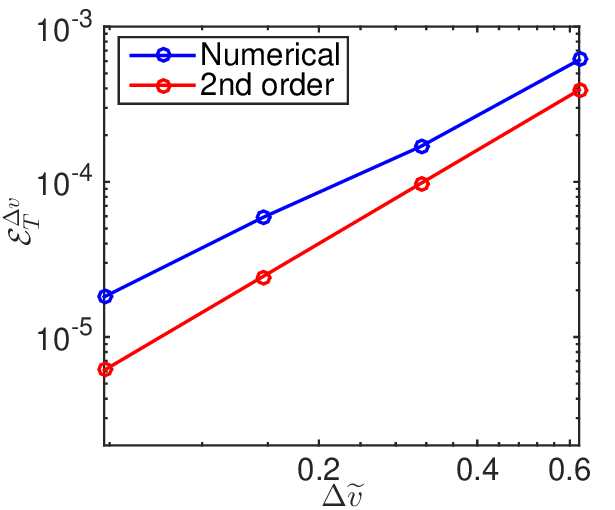}\includegraphics[width=5cm]{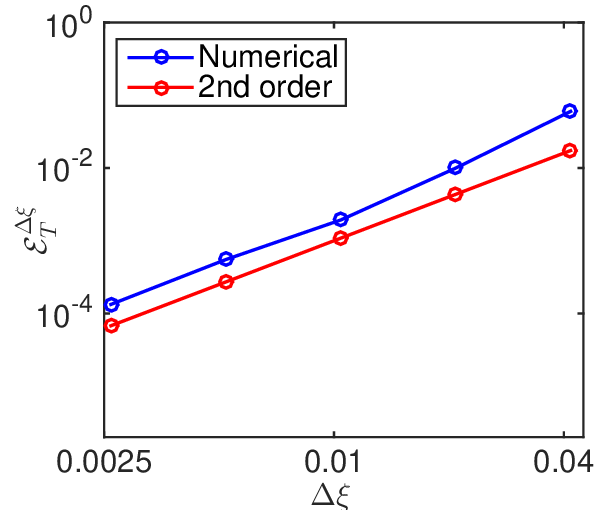}
\par\end{centering}
\caption{1D2V Sinusoidal perturbation relaxation: Temporal- (left), velocity-space-
(center), and spatial- (right) order accuracy of the proposed scheme.
Second-order accuracy is confirmed in all cases. \label{fig:1D2V_Sinusoidal_wave_relaxation_order_accuracy}}
\end{figure}

To demonstrate the second-order temporal convergence of the BDF2 scheme
that we employ, we compute the relative difference of the plasma temperature
with respect to the reference quantity,

\begin{equation}
{\cal E}_{T}^{\Delta t}=\sum_{i=1}^{N_{\xi}}\Delta\xi\frac{\left|T_{i}^{\Delta t,ref}-T_{i}^{\Delta t}\right|}{T_{i}^{\Delta t,ref}}.\label{eq:temporal_convergence_equation}
\end{equation}
Here, $T^{\Delta t,ref}$ is the reference temperature obtained using
a small time-step size ($\Delta t_{ref}=10^{-3}$) at the final time
$t_{max}=0.5$. For all cases, we use a grid size of $N_{\xi}=24$
and $N_{v}=32\times16$, and a solver nonlinear convergence tolerance
of of $\epsilon_{r}=10^{-6}$ (to adequately capture small signals
for small $\Delta t$). Fig. \ref{fig:1D2V_Sinusoidal_wave_relaxation_order_accuracy}-left
shows the expected asymptotic second order accuracy with $\Delta t$
refinement.

A second-order accuracy in velocity-space is demonstrated similarly
by computing:
\begin{equation}
{\cal E}_{T}^{\Delta\widetilde{v}}=\sum_{i=1}^{N_{\xi}}\Delta\xi\frac{\left|T_{i}^{\Delta\widetilde{v},ref}-T_{i}^{\Delta\widetilde{v}}\right|}{T_{i}^{\Delta\widetilde{v},ref}}.\label{eq:velocity_convergence_equation}
\end{equation}
Here, $T^{\Delta\widetilde{v},ref}$ is the reference temperature
solution obtained using a reference-grid resolution of $N_{v}^{ref}=512\times256$.
A uniform grid refinement is performed in both velocity-space directions.
For all cases, we use $\Delta t=5\times10^{-2}$ and a final time
$t_{max}=0.5$ with $N_{\xi}=24$. Fig. \ref{fig:1D2V_Sinusoidal_wave_relaxation_order_accuracy}-center
confirms the second-order convergence with $\Delta\widetilde{v}$
refinement. 

Finally, to demonstrate a second-order accuracy of the spatial discretization,
we use a similar approach and compute
\begin{equation}
{\cal E}_{T}^{\Delta\xi}=\sum_{i=1}^{N_{\xi,ref}}\Delta\xi_{ref}\frac{\left|T_{i}^{\Delta\xi,ref}-T_{i}^{\Delta\xi}\right|}{T_{i}^{\Delta\xi,ref}}.\label{eq:spatial_convergence_equation}
\end{equation}
Here, $T^{\Delta\xi,ref}$ and $\Delta\xi_{ref}$ are the reference-temperature
solution and logical-grid size, respectively, obtained using a reference-grid
resolution ($N_{\xi,ref}=768$) with $\Delta t=5\times10^{-2}$ and
$t_{max}=0.5$. To compute the norm in Eq. (\ref{eq:spatial_convergence_equation}),
we interpolate the coarse solution onto the reference configuration-space
grid (e.g., in $x$) via a $4^{th}$ order spline. For all cases,
we use a velocity space grid size of $N_{v}=32\times16$. Fig. \ref{fig:1D2V_Sinusoidal_wave_relaxation_order_accuracy}-right
confirms the expected second order of accuracy of our spatial discretization.

\subsection{1D2V Mach 5 Standing Shock}

We simulate a Mach-5 standing shock problem. The purpose of this simulation
is to demonstrate that the proposed phase-space grid adaptivity and
associated discrete-conservation strategy can reproduce published
results in the literature \citep{Taitano_2018_vrfp_1d2v_implicit,Vidal_PoP_1993_ion_kin_sim_of_plan_col_shock}.
We obtain the hydrodynamic jump conditions from the Hugoniot relationship:

\begin{equation}
\frac{P_{1}}{P_{0}}=\frac{2\gamma M^{2}-(\gamma-1)}{\gamma+1},
\end{equation}

\begin{equation}
\frac{\rho_{1}}{\rho_{0}}=\frac{u_{0}}{u_{1}}=\frac{M^{2}\left(\gamma+1\right)}{M^{2}\left(\gamma-1\right)+2}.
\end{equation}
Here, the subscript $0$ denotes the upstream (un-shocked) region
and the subscript $1$ denotes the downstream (shocked) region. Combining
these equations gives:

\begin{equation}
\frac{u_{0}}{u_{1}}=\frac{\left(\gamma-1\right)P_{0}+\left(\gamma+1\right)P_{1}}{\left(\gamma+1\right)P_{0}+\left(\gamma-1\right)P_{1}}.
\end{equation}
Here, $\gamma$ is the specific heat ratio ($\gamma=5/3$ for fully
ionized plasmas), $P$ is the total plasma pressure (i.e., $P=P_{i}+P_{e}$),
$\rho=\sum_{\alpha=1}^{N_{s}}m_{\alpha}n_{\alpha}$ is the total mass
density, and $u=\sum_{\alpha=1}^{N_{s}}m_{\alpha}n_{\alpha}u_{\alpha}/\sum_{\alpha=1}^{N_{s}}m_{\alpha}n_{\alpha}$
is the mass averaged drift velocity of the respective regions. The
upstream velocity can be expressed as

\begin{equation}
u_{0}=Mc_{0},
\end{equation}
where $c_{0}$ is the upstream sound speed,

\begin{equation}
c_{0}=\sqrt{\gamma\frac{P_{0}}{\rho_{0}}}.
\end{equation}
Employing the downstream conditions of $\rho_{1}=mn_{1}=1,$ $n_{1}=1$,
$m_{1}=1$, $P_{1}=P_{1,i}+P_{1,e}=2P_{1,i}=2n_{1,i}T_{1,i}=2$, $T_{1,i}=1$,
and $M=5$ gives for the upstrea\textcolor{black}{m conditions $\rho_{0}=mn_{0}=0.28$,
$P_{0}=0.1290$, and $T_{0}=0.1152$ . Then, $c_{0}=0.6197$, $u_{0}=Mc_{0}=3.0984$,
and $u_{1}=0.8676$. We note that, for this case and for comparison
purposes with Ref. \citep{Taitano_2018_vrfp_1d2v_implicit}, we employ
a variable Coulomb logarithm for both ion-ion and ion-electron collisions
\citep{plasma-formulary}, with $T^{0}=10$ keV and $n^{0}=10^{22}$
cm$^{-3}$. }

\textcolor{black}{To stress the computational savings afforded by
the adaptivity strategy, we consider a computational domain of $x\times\widetilde{v}_{||}\times\widetilde{v}_{\perp}\in\left[0,200\right]\times\left[-6,6\right]\times\left[0,6\right]$,
with a grid of $N_{\xi}=48$, $N_{v}=64\times32$. For comparison,
Ref. \citep{Taitano_2018_vrfp_1d2v_implicit} used $N_{x}=192$, $N_{v}=128\times64$.
The solution is initialized with a smoothed hyperbolic tangent with
the following profiles:}

\textcolor{black}{
\[
n=a_{n}\textnormal{tanh}\left[\mu\left(x-100\right)\right]+b_{n},\;u=a_{u}\textnormal{tanh}\left[\mu\left(x-100\right)\right]+b_{u},\;T=a_{T}\textnormal{tanh}\left[\mu\left(x-100\right)\right]+b_{T}
\]
where $a_{n}=0.36$, $b_{n}=0.64$, $a_{u}=-1.1154$, $b_{u}=1.983$,
$a_{T}=0.4424$, $b_{T}=0.5576$, and $\mu=0.05$. }

\textcolor{black}{In the configuration space, we consider the following
in- and out-flow boundary conditions for the ion distribution functions:}

\textcolor{black}{
\begin{equation}
f_{B}\left(v_{||},v_{\perp}\right)=\begin{cases}
f_{M}\left(n_{B},u_{B},T_{B}\right) & \textnormal{if}\;\widehat{l}_{B}v_{||}\le0\\
f_{C} & \textnormal{otherwise}
\end{cases}.
\end{equation}
Here, $n_{B}$, $u_{B}$, and $T_{B}$ are the moments defined by
the Hugoniot conditions at the boundary, $\widehat{l}_{B}$ is the
$x$-component of the boundary surface normal vector ($\pm1$ in 1D),
and $f_{C}$ is the distribution function defined in the computational
cell inside the domain, adjacent to the boundary. For the fluid-electron
temperature, we use the Dirichlet boundary conditions to impose the
Hugoniot asymptotic jump. }

\textcolor{black}{The simulation was run to $t_{max}=250$ until transient
structures had equilibrated. }
\begin{figure}[t]
\begin{centering}
\includegraphics[height=6cm]{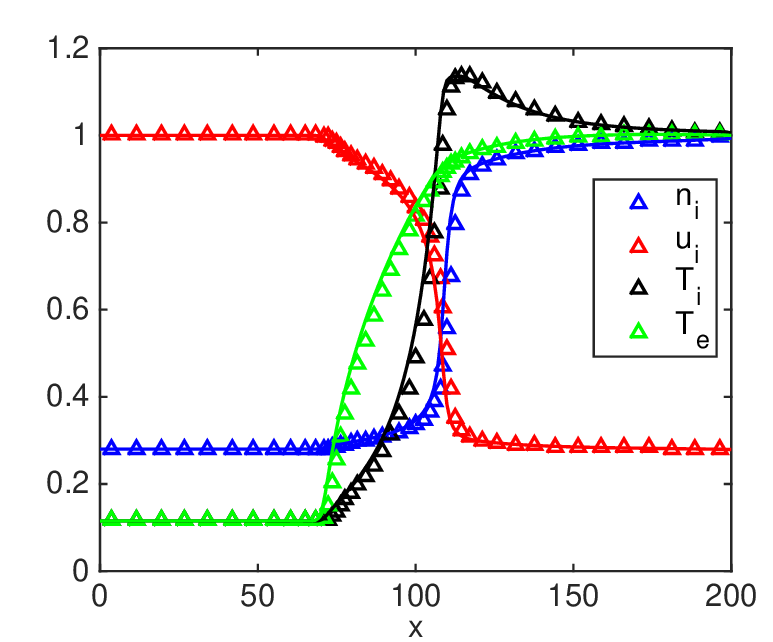}\includegraphics[height=6cm]{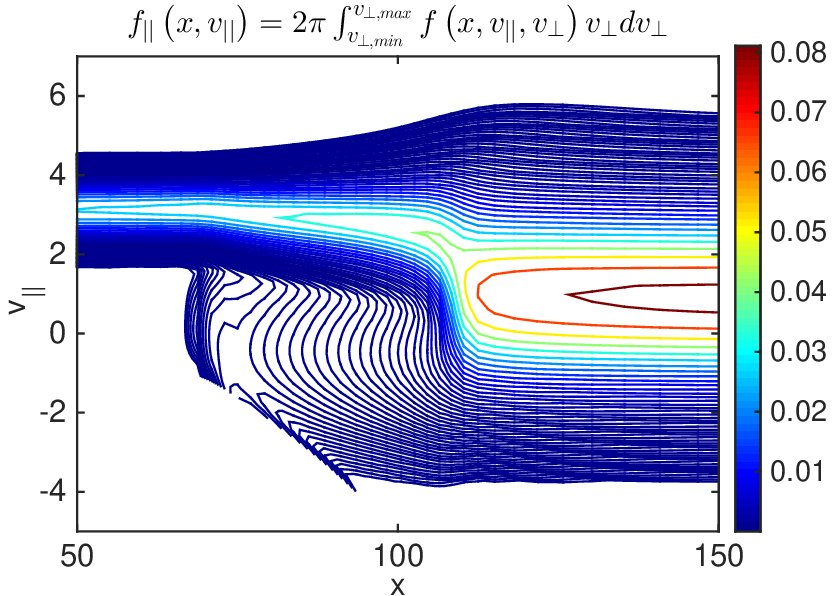}
\par\end{centering}
\textcolor{black}{\caption{Mach 5 steady-state shock: Left--The triangle markers are from our
simulation, while the solid lines are from Ref. \citep{Taitano_2018_vrfp_1d2v_implicit}
for a similar setup. We note that in the plot, the drift velocity,
$u$, is normalized to the upstream value. \textcolor{black}{Right--
The perpendicular-velocity-integrated distribution function of the
protons.}\textcolor{blue}{{} \label{fig:M=00003D5_standing_shock_plot}}}
}
\end{figure}
\textcolor{black}{Excellent agreement is found with respect to the
reference solution \citep{Taitano_2018_vrfp_1d2v_implicit}; refer
to Fig. \ref{fig:M=00003D5_standing_shock_plot}-left. Also seen in
Fig. \ref{fig:M=00003D5_standing_shock_plot}-right is a highly non-Maxwellian
feature in the distribution function, similar to those observed in
Fig. 4.8 of Ref. \citep{Taitano_2018_vrfp_1d2v_implicit}. In Fig.
\ref{fig:1d2v_m=00003D5_grid_and_temperature}, we also show the evolution
of the configuration-space grids at different times to demonstrate
the MMPDE strategy based on the inverse-gradient-length scale monitor
function in Eq. (\ref{eq:inverse_grad_scale_length_monitor_function}).
}
\begin{figure}[t]
\begin{centering}
\includegraphics[height=5cm]{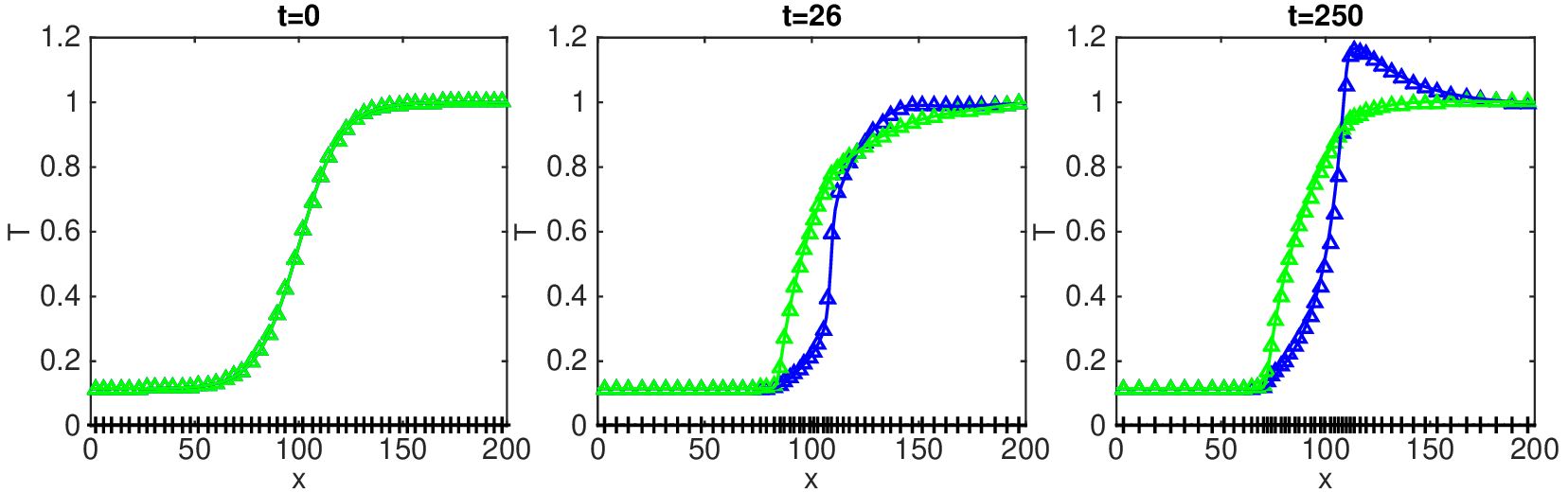}
\par\end{centering}
\caption{Mach 5 steady-state shock: Proton (blue) and electron (green) temperatures
at initial (left), intermediate (center), and steady-state (right)
times. The solid lines denote the solution obtained by a uniform-static
mesh in configuration space ($N_{\xi}=192$) while the triangle markers
denote the solution obtained with a coarse grid and adaptivity. As
can be seen, the configuration-space grid (+) adapts to the variation
in the gradients of temperature, as well as density and drift velocity
(not shown).\label{fig:1d2v_m=00003D5_grid_and_temperature}}
\end{figure}
As can be seen, the configuration-space grid tracks evolving features
near sharp gradients 1) without changing the total number of unknowns
as other adaptivity strategies do (e.g., AMR) and 2) without significantly
compromising the quality of the solution.

\begin{figure}[t]
\centering{}\includegraphics[height=4.8cm]{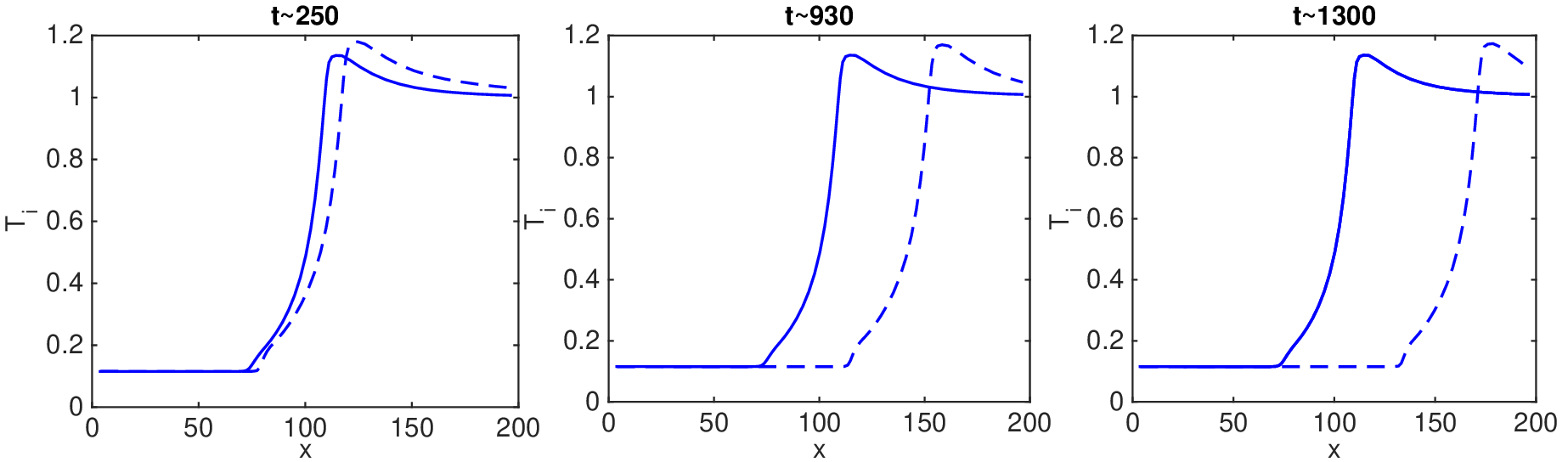}\caption{Mach 5 steady-state shock: Comparison of proton temperatures obtained
with (solid) and without (dashed) enforcing the Vlasov discrete conservation
symmetries.\label{fig:1d2v_m=00003D5_conservation_violation_consequence}}
\end{figure}

Finally, to stress once again the importance of enforcing discrete
conservation, we compare the ion temperature of the simulations with
and without the Vlasov temporal- and spatial-conservation symmetries
enforced; refer to Fig. \ref{fig:1d2v_m=00003D5_conservation_violation_consequence}.
As can be seen, without discrete conservation, the shock front drifts
indefinitely, resulting in ${\cal O}\left(1\right)$ error. In multiple
time-scale simulations, such subtle features can manifest as a departure
of the solution from the asymptotic slow manifold (e.g., hydrodynamic
limit), polluting the numerical results and making their physical
interpretation difficult.

\section{Conclusion\label{sec:conclusions}}

In this study, we have demonstrated, for the first time, an approach
that is fully conservative and adaptive in the phase space $\left(x,\vec{v}\right)$
for the multi-species, 1D2V VFP ion plasma equations with fluid electrons.
The approach features exact (in practice, up to a nonlinear tolerance)
mass, momentum, and energy conservation. Our approach analytically
adapts the velocity-space mesh for each species by normalizing the
velocity space to each species' reference speed, $v^{*}$, and shifting
by a translation velocity, $u_{||}^{*}$ (i.e., we consider multiple
velocity-space grids), and allows for an arbitrary variation (in practice,
to acceptable level of truncation error) in $v_{th}$ and $u_{||}$.
In configuration-space, we transform the VFP-ion and fluid-electron
equations on a logically Cartesian grid, and the Jacobian of transformation
is evolved using an MMPDE formalism. The analytical formulation allows
us to expose the continuum-conservation symmetries in the inertial
terms arising from the coordinate transformation, which are then enforced
discretely via the use of discrete-nonlinear constraint functions,
as proposed in earlier studies \citep{Taitano_2015_jcp_cmec_va,Taitano_2015_rfp_0d2v_implicit,Taitano_2016_rfp_0d2v_implicit,Taitano_2018_vrfp_1d2v_implicit}.
\textcolor{black}{We have shown the importance of enforcing the discrete
conservation and the adverse effects of not doing so for several test
cases. We close by noting that the methodology developed in this study
has been extended to a spherical geometry with an imploding boundary
to study ICF capsule implosion \citep{taitano_pop_2018}. This work
will be documented in a follow-on manuscript.}

\section*{Acknowledgments}

This work was sponsored by the Metropolis Postdoctoral Fellowship
for W.T.T. between the years 2015-2017, the LDRD office between the
years 2017-2018, the Institutional Computing program, and the Thermonuclear
Burn Initiative of the Advanced Simulation and Computing Program at
the Los Alamos National Laboratory. This work was performed under
the auspices of the National Nuclear Security Administration of the
U.S. Department of Energy at Los Alamos National Laboratory, managed
by Triad National Security, LLC under contract 89233218CNA000001.

\bibliographystyle{ieeetr}
\bibliography{./mybib,./kinetic,./fokker-planck,./transport,./numerics,./general,./icf_physics,./mmpde,./phase_space_adaptivity}

\appendix

\section{Derivation of Vlasov-Fokker-Planck Equation in the Transformed Coordinate
System\label{app:coordinate_transformation_details}}

Beginning with the 1D Vlasov equation in the original coordinates,
$\left(x,\vec{v},t\right)$,

\begin{equation}
\left.\frac{\partial f}{\partial t}\right|_{x,\vec{v}}+v_{||}\left.\frac{\partial f}{\partial x}\right|_{\vec{v},t}+\frac{q}{m}E_{||}\left.\frac{\partial f}{\partial v_{||}}\right|_{x,t}=0,\label{eq:app_vlasov}
\end{equation}
and introducing the coordinates, $\left(\xi,\vec{\widetilde{v}},t\right)$,
we expand each term as follows:

\begin{eqnarray}
\left.\frac{\partial f}{\partial t}\right|_{x,\vec{v}}=\left.\frac{\partial f}{\partial t}\right|_{\xi,\vec{v}}-\left.\frac{\partial f}{\partial\xi}\right|_{\vec{v},t}J_{\xi}^{-1}\dot{x}\nonumber \\
=\left.\frac{\partial f}{\partial t}\right|_{\xi,\vec{\widetilde{v}}}+\left.\frac{\partial f}{\partial\widetilde{\vec{v}}}\right|_{\xi,t}\cdot\left(\frac{\partial\widetilde{\vec{v}}}{\partial v^{*}}\frac{\partial v^{*}}{\partial t}+\frac{\partial\widetilde{\vec{v}}}{\partial\widehat{\vec{u}}^{*}}\cdot\frac{\partial\widehat{\vec{u}}^{*}}{\partial t}\right)-\left[\left.\frac{\partial f}{\partial\xi}\right|_{\widetilde{\vec{v}},t}+\left.\frac{\partial f}{\partial\widetilde{\vec{v}}}\right|_{\xi,t}\cdot\left(\frac{\partial\widetilde{\vec{v}}}{\partial v^{*}}\frac{\partial v^{*}}{\partial\xi}+\frac{\partial\widetilde{\vec{v}}}{\partial\widehat{\vec{u}}^{*}}\cdot\frac{\partial\widehat{\vec{u}}^{*}}{\partial\xi}\right)\right]J_{\xi}^{-1}\dot{x},\label{eq:app_temporal_expansion}
\end{eqnarray}

\begin{eqnarray}
v_{||}\left.\frac{\partial f}{\partial x}\right|_{\vec{v},t}=v^{*}\left(\widetilde{v}_{||}+\widehat{u}_{||}^{*}\right)\left.\frac{\partial f}{\partial\xi}\right|_{\vec{v},t}J^{-1}=v^{*}\left(\widetilde{v}_{||}+\widehat{u}_{||}^{*}\right)\left[\left.\frac{\partial f}{\partial\xi}\right|_{\widetilde{\vec{v}},t}+\left.\frac{\partial f}{\partial\widetilde{\vec{v}}}\right|_{\xi,t}\cdot\left(\frac{\partial\widetilde{\vec{v}}}{\partial v^{*}}\frac{\partial v^{*}}{\partial\xi}+\frac{\partial\widetilde{\vec{v}}}{\partial\widehat{\vec{u}}^{*}}\cdot\frac{\partial\widehat{\vec{u}}^{*}}{\partial\xi}\right)\right]J_{\xi}^{-1},\label{eq:app_spatial_expansion}
\end{eqnarray}
and

\begin{equation}
\frac{q}{m}E_{||}\left.\frac{\partial f}{\partial v_{||}}\right|_{x,t}=\frac{q}{m}\frac{E_{||}}{v^{*}}\left.\frac{\partial f}{\partial\widetilde{v}_{||}}\right|_{\xi,t},\label{eq:app_acceleration_expansion}
\end{equation}
where, 
\[
J_{\xi}^{-1}=\frac{\partial\xi}{\partial x},\;\dot{x}=\frac{\partial x}{\partial t},\frac{\partial\widetilde{\vec{v}}}{\partial v^{*}}=-\frac{\widetilde{\vec{v}}+\widehat{\vec{u}}^{*}}{v^{*}},\;\frac{\partial\widetilde{\vec{v}}}{\partial\vec{\widehat{u}}^{*}}=-\tensor I,\;\textnormal{and}\;\widehat{\vec{u}}^{*}=\widehat{u}_{||}^{*}\vec{e}_{||}.
\]
Assembling the terms, we obtain:

\begin{eqnarray}
\left.\frac{\partial f}{\partial t}\right|_{\xi,\vec{\widetilde{v}}}-\left.\frac{\partial f}{\partial\widetilde{\vec{v}}}\right|\cdot\left(\frac{\widetilde{\vec{v}}+\widehat{u}_{||}^{*}\vec{e}_{||}}{v^{*}}\frac{\partial v^{*}}{\partial t}+\frac{\partial\widehat{u}_{||}^{*}}{\partial t}\vec{e}_{||}\right)-\left[\left.\frac{\partial f}{\partial\xi}\right|_{\widetilde{\vec{v}},t}-\left.\frac{\partial f}{\partial\widetilde{\vec{v}}}\right|_{\xi,t}\cdot\left(\frac{\widetilde{\vec{v}}+\widehat{u}_{||}^{*}\vec{e}_{||}}{v^{*}}\frac{\partial v^{*}}{\partial\xi}+\frac{\partial\widehat{u}_{||}^{*}}{\partial\xi}\vec{e}_{||}\right)\right]J_{\xi}^{-1}\dot{x},\nonumber \\
+v^{*}\left(\widetilde{v}_{||}+\widehat{u}_{||}^{*}\right)\left[\left.\frac{\partial f}{\partial\xi}\right|_{\widetilde{\vec{v}},t}-\left.\frac{\partial f}{\partial\widetilde{\vec{v}}}\right|_{\xi,t}\cdot\left(\frac{\widetilde{\vec{v}}+\widehat{u}_{||}^{*}\vec{e}_{||}}{v^{*}}\frac{\partial v^{*}}{\partial\xi}+\frac{\partial\widehat{u}_{||}^{*}}{\partial\xi}\vec{e}_{||}\right)\right]J_{\xi}^{-1}+\frac{q}{m}\frac{E_{||}}{v^{*}}\left.\frac{\partial f}{\partial\widetilde{v}_{||}}\right|_{\xi,t}=0.\label{eq:app_expanded_vlasov}
\end{eqnarray}

Noting that $f=\frac{\widetilde{f}}{\left(v^{*}\right)^{3}}$,

\[
\left.\frac{\partial f}{\partial t}\right|_{\xi,\widetilde{\vec{v}}}=\frac{1}{\left(v^{*}\right)^{3}}\left.\frac{\partial\widetilde{f}}{\partial t}\right|_{\xi,\widetilde{\vec{v}}}-\frac{3\widetilde{f}}{\left(v^{*}\right)^{4}}\frac{\partial v^{*}}{\partial t},
\]
\[
\left.\frac{\partial f}{\partial\xi}\right|_{\widetilde{\vec{v}},t}=\frac{1}{\left(v^{*}\right)^{3}}\left.\frac{\partial\widetilde{f}}{\partial\xi}\right|_{\widetilde{\vec{v}},t}-\frac{3\widetilde{f}}{\left(v^{*}\right)^{4}}\frac{\partial v^{*}}{\partial\xi},
\]
and multiplying Eq. (\ref{eq:app_expanded_vlasov}) by $\left(v^{*}\right)^{3}J_{\xi}$,
one obtains,

\begin{eqnarray}
J_{\xi}\left.\frac{\partial\widetilde{f}}{\partial t}\right|_{\xi,\vec{\widetilde{v}}}-J_{\xi}\frac{3\widetilde{f}}{v^{*}}\frac{\partial v^{*}}{\partial t}-J_{\xi}\left.\frac{\partial\widetilde{f}}{\partial\widetilde{\vec{v}}}\right|\cdot\left(\frac{\widetilde{\vec{v}}+\widehat{u}_{||}^{*}\vec{e}_{||}}{v^{*}}\frac{\partial v^{*}}{\partial t}+\frac{\partial\widehat{u}_{||}^{*}}{\partial t}\vec{e}_{||}\right)\nonumber \\
-\left[\left.\frac{\partial\widetilde{f}}{\partial\xi}\right|_{\widetilde{\vec{v}},t}-\frac{3\widetilde{f}}{v^{*}}\frac{\partial v^{*}}{\partial\xi}-\left.\frac{\partial\widetilde{f}}{\partial\widetilde{\vec{v}}}\right|_{\xi,t}\cdot\left(\frac{\widetilde{\vec{v}}+\widehat{u}_{||}^{*}\vec{e}_{||}}{v^{*}}\frac{\partial v^{*}}{\partial\xi}+\frac{\partial\widehat{u}_{||}^{*}}{\partial\xi}\vec{e}_{||}\right)\right]\dot{x}\nonumber \\
+v^{*}\left(\widetilde{v}_{||}+\widehat{u}_{||}^{*}\right)\left[\left.\frac{\partial\widetilde{f}}{\partial\xi}\right|_{\widetilde{\vec{v}},t}-\frac{3\widetilde{f}}{v^{*}}\frac{\partial v^{*}}{\partial\xi}-\left.\frac{\partial\widetilde{f}}{\partial\widetilde{\vec{v}}}\right|_{\xi,t}\cdot\left(\frac{\widetilde{\vec{v}}+\widehat{u}_{||}^{*}\vec{e}_{||}}{v^{*}}\frac{\partial v^{*}}{\partial\xi}+\frac{\partial\widehat{u}_{||}^{*}}{\partial\xi}\vec{e}_{||}\right)\right]+J_{\xi}\frac{q}{m}\frac{E}{v^{*}}\left.\frac{\partial\widetilde{f}}{\partial\widetilde{v}_{||}}\right|_{\xi,t}=0.\label{eq:expanded_vlasov_tilde}
\end{eqnarray}
Recognizing the following chain-rules,

\[
\frac{\partial}{\partial\vec{\widetilde{v}}}\cdot\left\{ \left(\frac{\widetilde{\vec{v}}+\widehat{u}_{||}^{*}\vec{e}_{||}}{v^{*}}\frac{\partial v^{*}}{\partial t}+\frac{\partial\widehat{u}_{||}^{*}}{\partial t}\vec{e}_{||}\right)\widetilde{f}\right\} _{\xi,t}=\frac{3\widetilde{f}}{v^{*}}\frac{\partial v^{*}}{\partial t}+\left.\frac{\partial\widetilde{f}}{\partial\widetilde{\vec{v}}}\right|_{\xi,t}\cdot\left(\frac{\widetilde{\vec{v}}+\widehat{u}_{||}^{*}\vec{e}_{||}}{v^{*}}\frac{\partial v^{*}}{\partial t}+\frac{\partial\widehat{u}_{||}^{*}}{\partial t}\vec{e}_{||}\right),
\]

\begin{eqnarray*}
\frac{\partial}{\partial\vec{\widetilde{v}}}\cdot\left\{ \left[v^{*}\left(\widetilde{v}_{||}+\widehat{u}_{||}^{*}\right)-\dot{x}\right]\left(\frac{\widetilde{\vec{v}}+\widehat{u}_{||}^{*}\vec{e}_{||}}{v^{*}}\frac{\partial v^{*}}{\partial\xi}+\frac{\partial\widehat{u}_{||}^{*}}{\partial\xi}\vec{e}_{||}\right)\widetilde{f}\right\} _{\xi,t}=\\
3\widetilde{f}\left\{ \frac{\left[v^{*}\left(\widetilde{v}_{||}+\widehat{u}_{||}^{*}\right)-\dot{x}\right]}{v^{*}}\frac{\partial v^{*}}{\partial\xi}+\left[\frac{\widetilde{v}_{||}+\widehat{u}_{||}^{*}}{v^{*}}\frac{\partial v^{*}}{\partial\xi}+\frac{\partial\widehat{u}_{||}^{*}}{\partial\xi}\right]v^{*}\right\} \\
+\left.\frac{\partial\widetilde{f}}{\partial\widetilde{\vec{v}}}\right|_{\xi,t}\cdot\left\{ \left[v^{*}\left(\widetilde{v}_{||}+\widehat{u}_{||}^{*}\right)-\dot{x}\right]\left(\frac{\widetilde{\vec{v}}+\widehat{u}_{||}^{*}\vec{e}_{||}}{v^{*}}\frac{\partial v^{*}}{\partial\xi}+\frac{\partial\widehat{u}_{||}^{*}}{\partial\xi}\vec{e}_{||}\right)\right\} ,
\end{eqnarray*}

\[
\left.\frac{\partial}{\partial t}\left(J_{\xi}\widetilde{f}\right)\right|_{\xi,\widetilde{\vec{v}}}=J_{\xi}\left.\frac{\partial\widetilde{f}}{\partial t}\right|_{\xi,\widetilde{\vec{v}}}+\widetilde{f}\frac{\partial\dot{x}}{\partial\xi},
\]
\[
\left.\frac{\partial}{\partial\xi}\left(\dot{x}\widetilde{f}\right)\right|_{\widetilde{\vec{v}},t}=\dot{x}\left.\frac{\partial\widetilde{f}}{\partial\xi}\right|_{\widetilde{\vec{v}},t}+\widetilde{f}\frac{\partial\dot{x}}{\partial\xi},
\]
\[
\frac{\partial}{\partial\xi}\left[v_{||}^{*}\left(\widetilde{v}_{||}+\widehat{u}_{||}^{*}\right)\widetilde{f}\right]_{\widetilde{\vec{v}},t}=v^{*}\left(\widetilde{v}_{||}+\widehat{u}_{||}^{*}\right)\left.\frac{\partial\widetilde{f}}{\partial\xi}\right|_{\widetilde{\vec{v}},t}+\widetilde{f}v^{*}\left(\frac{\widetilde{v}_{||}+\widehat{u}_{||}^{*}}{v^{*}}\frac{\partial v^{*}}{\partial\xi}+\frac{\partial\widehat{u}_{||}^{*}}{\partial\xi}\right),
\]
\[
\frac{1}{v^{*}}\frac{\partial}{\partial t}\left[v^{*}\left(\vec{\widetilde{v}}+\widehat{u}_{||}^{*}\vec{e}_{||}\right)\right]=\frac{\widetilde{\vec{v}}+\widehat{u}_{||}^{*}\vec{e}_{||}}{v^{*}}\frac{\partial v^{*}}{\partial t}+\frac{\partial\widehat{u}_{||}^{*}}{\partial t}\vec{e}_{||},
\]
\[
\frac{1}{v^{*}}\frac{\partial}{\partial\xi}\left[v^{*}\left(\vec{\widetilde{v}}+\widehat{u}_{||}^{*}\vec{e}_{||}\right)\right]=\frac{\widetilde{\vec{v}}+\widehat{u}_{||}^{*}\vec{e}_{||}}{v^{*}}\frac{\partial v^{*}}{\partial\xi}+\frac{\partial\widehat{u}_{||}^{*}}{\partial\xi}\vec{e}_{||}.
\]
and substituting these results into the original Vlasov-Fokker-Planck
equation, Eq. (\ref{eq:1d_vfp_eqn}), there results the following
transformed equation for $\widetilde{f}_{\alpha}\left(\xi,\widetilde{\vec{v}},t\right)$,
\begin{eqnarray}
\left.\frac{\partial\left(J_{\xi}\widetilde{f}_{\alpha}\right)}{\partial t}\right|_{\xi,\widetilde{\vec{v}}}+\frac{\partial}{\partial\xi}\left[\left(v_{\alpha}^{*}\left[\widetilde{v}_{||}+\widehat{u}_{||,\alpha}^{*}\right]-\dot{x}\right)\widetilde{f}_{\alpha}\right]_{\widetilde{\vec{v}},t}-\frac{J_{\xi}}{v_{\alpha}^{*}}\frac{\partial}{\partial\widetilde{\vec{v}}}\cdot\left[\frac{\partial}{\partial t}\left(v_{\alpha}^{*}\left[\vec{\widetilde{v}}+\widehat{u}_{||,\alpha}^{*}\vec{e}_{||}\right]\right)\widetilde{f}_{\alpha}\right]_{\xi,t}\nonumber \\
-\frac{1}{v_{\alpha}^{*}}\frac{\partial}{\partial\widetilde{\vec{v}}}\cdot\left[\frac{\partial}{\partial\xi}\left(v_{\alpha}^{*}\left[\vec{\widetilde{v}}+\widehat{u}_{||,\alpha}^{*}\vec{e}_{||}\right]\right)\left(v_{\alpha}^{*}\left[\widetilde{v}_{||}+\widehat{u}_{||,\alpha}^{*}\right]-\dot{x}\right)\widetilde{f}_{\alpha}\right]_{\xi,t}+J_{\xi}\frac{q_{\alpha}}{m_{\alpha}}\frac{E_{||}}{v_{\alpha}^{*}}\left.\frac{\partial\widetilde{f}_{\alpha}}{\partial\widetilde{v}_{||}}\right|_{\xi,t}=J_{\xi}\left[\sum_{\beta}\widetilde{C}_{\alpha\beta}+\widetilde{C}_{\alpha e}\right],\label{eq:vfp_eqn_transformed_2}
\end{eqnarray}
where we have used the definition of the normalized collision operator,
$\widetilde{C}_{\alpha\beta}=\left(v_{\alpha}^{*}\right)^{3}C_{\alpha\beta}$
(for further details, we refer the readers to Ref. \citep{Taitano_2016_rfp_0d2v_implicit}). 
\end{document}